\documentclass{pasa}%

\usepackage{graphicx}
\usepackage{rotating}
\usepackage{lscape}
\usepackage{morefloats}
\usepackage{amssymb}
\usepackage{url}
\usepackage{natbib}

\usepackage{aas_macros}
\usepackage{hyperref}
\hypersetup{colorlinks,citecolor=blue,linkcolor=blue,urlcolor=blue}

\usepackage{geometry}

\def\simlt{\mathrel{\rlap{\lower 3pt\hbox{$\sim$}}\raise 2.0pt\hbox{$<$}}}
\def\simgt{\mathrel{\rlap{\lower 3pt\hbox{$\sim$}} \raise
2.0pt\hbox{$>$}}}
\def\lsim{\,\lower2truept\hbox{${<\atop\hbox{\raise4truept\hbox{$\sim$}}}$}\,}
\def\gsim{\,\lower2truept\hbox{${> \atop\hbox{\raise4truept\hbox{$\sim$}}}$}\,}

\DeclareMathAlphabet{\mathpzc}{OT1}{pzc}{m}{it}

\begin{document}

\title[All-sky submm spectroscopic surveys]{Understanding galaxy formation and evolution through an all-sky submillimetre spectroscopic survey}


\author[Negrello et al.]
{Mattia Negrello$^1$, Matteo Bonato$^{2,3}$, Zhen-Yi Cai$^{4,5}$, Helmut
Dannerbauer$^{6,7}$\thanks{helmut@iac.es, corresponding author}, Gianfranco De
Zotti$^3$\thanks{gianfranco.dezotti@inaf.it, corresponding author}, Jacques
Delabrouille$^{8,9}$, Douglas Scott$^{10}$
\affil{$^1$School of Physics and Astronomy, Cardiff University, The Parade, Cardiff CF24 3AA, UK}%
\affil{$^2$INAF$-$Istituto di Radioastronomia, and Italian ALMA Regional Centre, Via Gobetti 101, I-40129, Bologna, Italy}%
\affil{$^3$INAF, Osservatorio Astronomico di Padova, Vicolo Osservatorio 5, I-35122 Padova, Italy}%
\affil{$^4$CAS Key Laboratory for Research in Galaxies and Cosmology,
Department of Astronomy, University of Science and Technology of China, Hefei 230026, China}%
\affil{$^5$School of Astronomy and Space Science, University of Science and
Technology of China, Hefei 230026, China}%
\affil{$^6$Instituto de Astrof{\'\i}sica de Canarias (IAC), E-38205 La Laguna, Tenerife, Spain}%
\affil{$^7$Universidad de La Laguna, Dpto.  Astrof{\'\i}sica, E-38206 La Laguna, Tenerife, Spain}%
\affil{$^8$Laboratoire Astroparticule et Cosmologie, CNRS/IN2P3, 75205 Paris Cedex 13, France}%
\affil{$^9$D\'epartement d'Astrophysique, CEA Saclay DSM/Irfu, 91191 Gif-sur-Yvette, France}%
\affil{$^{10}$Department of Physics \& Astronomy,  University of British
Columbia, Vancouver,  Canada} }

\jid{PASA} \doi{10.1017/pas.\the\year.xxx} \jyear{\the\year}


\hypersetup{draft}


\begin{frontmatter}
\maketitle

\begin{abstract}
We illustrate the extraordinary discovery potential for extragalactic
astrophysics of a far-infrared/submillimetre (far-IR/submm) all-sky
spectroscopic survey with a 3-m-class space telescope. Spectroscopy provides
both a 3-dimensional view of the Universe and allows us to take full
advantage of the sensitivity of present-day instrumentation, close to
fundamental limits, overcoming the spatial  confusion that affects broadband
far-IR/submm surveys. A space telescope of the 3-m class (which has already
been described in recent papers) will detect emission lines powered by star
formation in galaxies out to $z\,{\simeq}\,8$. It will specifically provide
measurements of spectroscopic redshifts, star-formation rates, dust masses,
and metal content for millions of galaxies at the peak epoch of cosmic star
formation and of hundreds of them at the epoch of reionization. Many of these
star-forming galaxies will be strongly lensed; the brightness amplification
and stretching of their sizes will make it possible to investigate (by means
of follow-up observations with high-resolution instruments like ALMA, {\it
JWST}, and SKA) their internal structure and dynamics on the scales of giant
molecular clouds (40--100\,pc).  This will provide direct information on the
physics driving the evolution of star-forming galaxies. Furthermore, the
arc-minute resolution of the telescope at submm wavelengths is ideal for
detecting the cores of galaxy proto-clusters, out to the epoch of
reionization. Due to the integrated emission of member galaxies, such objects
(as well as strongly lensed sources) will dominate at the highest apparent
far-IR luminosities. Tens of millions of these galaxy-clusters-in-formation
will be detected at $z\,{\simeq}\,2$--3, with a tail extending out to
$z\,{\simeq}\,7$, and thousands of detections at $6\,{<}\,z\,{<}\,7$. Their
study will allow us to track the growth of the most massive halos well beyond
what is possible with classical cluster surveys (mostly limited to
$z\,{\simlt}\, 1.5$--2), tracing the history of star formation in dense
environments and teaching us how star formation and galaxy-cluster formation
are related across all epochs. The obscured cosmic star-formation-rate
density of the Universe will thereby be constrained. Such a survey will
overcome the current lack of spectroscopic redshifts of dusty star-forming
galaxies and galaxy proto-clusters, representing a quantum leap in
far-IR/submm extragalactic astrophysics.
\end{abstract}

\begin{keywords}
galaxies: luminosity function -- galaxies: evolution -- galaxies: clusters:
general -- galaxies: high-redshift -- submillimeter: galaxies
\end{keywords}
\end{frontmatter}




\linespread{0.94}
\voffset=-0.4cm






\synctex=1
\def\aap{A\&A}
\def\apj{ApJ}
\def\apjs{ApJS}
\def\apjl{ApJL}
\def\mnras{MNRAS}
\def\aj{AJ}
\def\nat{Nature}
\def\aaps{A\&A Supp.}
\def\pra{Phys. Rev. A}         
\def\prb{Phys. Rev. B}         
\def\prc{Phys. Rev. C}         
\def\prd{Phys. Rev. D}         
\def\prl{Phys. Rev. Lett}      
\def\araa{ARA\&A}       
\def\gca{GeCoA}         
\def\pasp{PASP}              
\def\pasj{PASJ}              
\def\apss{ApSS}
\def\jcap{JCAP}
\def\plb{Phys. Lett. B.}
\def\jhep{JHEP}
\def\physrep{Phys. Rep.}
\def\baas{BAAS}



\section{Introduction}
\label{sec:intro}

The L-class space mission proposed by \citet{Delabrouille2019} will have a
tremendous impact on our understanding of the Universe and on many branches of
astrophysics. The project features two instruments at the focus of a 3-m-class,
cold (8\,K) telescope: (i) a broad-band, multi-frequency, polarimetric imager
operating over the 20--800\,GHz frequency range that would map the cosmic
microwave background (CMB) at high sensitivity, as well as the thermal and
kinetic Sunyaev-Zeldovich (SZ) effects and the Galactic and extragalactic
continuum emissions; and (ii) a moderate spectral resolution
($R\,{\simeq}\,300$) filter-bank spectrometer covering the 100--1000\,GHz band.
These two instruments would comprise tens of thousands of millimetre (mm) and
submillimetre (submm) detectors cooled to sub-kelvin temperatures for
sky-background-limited performance.

A set of Fourier-transform spectrometers (FTSs), covering the full
10--2000\,GHz band with spectral resolution ranging from 2.5 to 60\,GHz, could
also be hosted on the same platform. This instrument would carry out absolute
measurements of the CMB spectrum with a sensitivity 4 to 5 orders of magnitude
better than {\it COBE}-FIRAS.

An overview of the scientific goals of the project has been  presented by
\citet{Delabrouille2019}, while other white papers related to the ESA ``Voyage
2050''\footnote{Call by ESA in order to prepare the long-term plan of the ESA
science program.} call have elaborated on specific science cases.
\citet{Basu2019} dealt with the use of the CMB as a ``back-light'',
illuminating the entire observable Universe, thus allowing us to obtain a
complete census of the total mass, gas, and stellar contents of the Cosmos
across time. \citet{Silva2019} looked into the promise of mapping the intensity
of the many mm/submm/far-IR lines detectable by the proposed instrument to
address several open questions relating to the reionization process, galaxy
evolution, the cosmic infrared background, and fundamental cosmology.
Additionally, the unique information on the thermal history of the Universe
provided by absolute spectral measurements has been highlighted by
\citet{Chluba2019}.

Here we argue that the high-sensitivity spectroscopic and imaging surveys
carried out by this space mission would revolutionize our understanding of
galaxy formation and evolution and of the growth of large-scale structure back
to the epoch of reionization.

For more than 20 year, surveys at far-infrared (FIR) to submm wavelengths have
played a key role in our understanding of early galaxy and active galactic
nuclei (AGN) evolution \citep[for reviews see][]{Lutz2014,Casey2014}. The SCUBA
discovery of a copious population of submm-bright galaxies \citep{Smail1997,
Barger1998, Hughes1998}, shown to be at high redshifts \citep{Ivison1998,
Barger1999, Swinbank2004, Chapman2005, Pope2006}, strongly challenged the
widely accepted pictures of galaxy formation and evolution. In fact, simple
merger-driven models dramatically underpredicted the abundance of bright submm
galaxies and the cosmic infrared background (CIB) intensity, using standard
assumptions for the stellar initial mass function (IMF) and for the dust
temperature distribution \citep[see, e.g.,][]{Kaviani2003, Baugh2005,
Somerville2012, Niemi2012, Gruppioni2015}.

The submm spectral region is exceptionally well suited to provide access to the
dust-enshrouded most active star-formation phases of young galaxies in the
high-$z$ Universe. This is because, for a large redshift range, submm
wavelengths (in the observer's frame) are in the Rayleigh-Jeans region of the
dust emission spectrum, where the flux density scales as $S_\nu
\propto\nu^{2+\beta}$, the dust emissivity index, $\beta$, generally being in
the range 1.5--2. The corresponding, so-called ``negative $K$-correction''
largely compensates, and may even slightly exceed, the effect on the flux
density of the increase of the luminosity distance \citep{Franceschini1991,
BlainLongair1993}, providing roughly luminosity-limited samples. Thus, dusty
star-forming galaxies (DSFGs) can be detected out to $z\,{\simeq}\,8$--10
without the need for extreme sensitivities.

Ground-based observations at submm wavelengths are severely limited by water
vapour in the atmosphere, leaving only a few windows even in the driest sites;
hence the need for space missions. The \textit{Herschel} observatory surveyed
about $1300\,\hbox{deg}^2$ of the extragalactic sky, primarily thanks to the
\textit{Herschel} Astrophysical Terahertz Large Area Survey
\citep[H-ATLAS;][]{Eales2010}, to the \textit{Herschel} Multi-tiered
Extragalactic Survey \citep[HerMES;][]{Oliver2012} and to the \textit{Herschel}
Stripe 82 Survey \citep[HerS;][]{Viero2014}, which covered $660\,\hbox{deg}^2$,
$380\,\hbox{deg}^2$ and $79\,\hbox{deg}^2$, respectively.

The \textit{Herschel} surveys with the Spectral and Photometric Imaging
Receiver (SPIRE), operating at 250, 350 and $500\,\mu$m, were confusion limited
at rms values of 5.8, 6.3, and $6.8\,{\rm mJy}\,{\rm beam}^{-1}$, respectively
\citep{Nguyen2010}. Source confusion does not allow us to take full advantage
of the sensitivity (which is close to fundamental limits) of present-day
detectors. The problem can be overcome using larger telescopes, thereby
providing better angular resolution, but this implies higher costs and limits
on the total area that can be surveyed.

An interesting alternative is offered by spectroscopy. Adding the third
dimension essentially removes the confusion problem. The reason for this is
clear -- moderate resolution spectroscopy with, say,
$R\,{\equiv}\,\nu/\Delta\nu\,{=}\,300$, apportions sources detected with
broadband photometry with, e.g., $\Delta\nu/\nu\,{=}\,0.3$ to 100 almost
equally populated narrow redshift bins, each with much lower, generally
negligible, confusion noise.  Precisely how to most efficiently disentangle the
spectra of sources from spatially-confused (but spectrally unconfused) data
cubes remains an open research question \cite[see discussion in appendix~E.1
of][]{Cooray2019}, but it is clear that in principle the information is there
to extract.

The relatively large beam sizes of submm telescopes have another serious
drawback: the difficulty in identifying multi-frequency counterparts to measure
their spectroscopic redshifts. High-$z$ dusty star-forming galaxies (DSFGs) are
generally very faint at other wavelengths, such as in the optical or near-IR
that do not benefit from the negative $K$-correction
\citep[e.g.,][]{Dannerbauer2002, Dannerbauer2004, Pope2005, Dannerbauer2008,
Dunlop2004, Younger2007, Walter2012}. The surface densities of faint
optical/near-IR sources are very high, making the identification of the correct
counterpart difficult. Deep radio observations help a lot in this respect
\citep[e.g.,][]{Ivison2002,Dannerbauer2004,Biggs2010}. Acquiring spectroscopic
redshifts through follow-up studies with mm/submm telescopes is a
time-consuming process, impractical for very large galaxy samples. Although
FIR/submm surveys have discovered tens of thousands of distant DSFGs, the
number of measured spectroscopic redshifts is still not more than a few
hundreds \citep[e.g.,][]{Casey2012a, Casey2012b, Bothwell2013, Weiss2013,
Danielson2017, Fudamoto2017, ZhangZ2018, Neri2019}. Thus, only coarse
FIR-photometric redshift estimates are generally possible for samples of
several hundred to thousands of high-$z$ IR-luminous objects.

A moderate spectral resolution ($R\,{\simeq}\,300$) filter-bank spectrometer
at the focus of a 3-m-class telescope will yield exciting results in several
areas. For definiteness we consider a cold (8\,K) telescope with a 3.5-m
aperture primary (the same size as \textit{Herschel}) with a secondary mirror
and cold stop at 4\,K and 20\,dB edge taper, as proposed by
\citet{Delabrouille2019}.

\begin{figure*}[htbp!]
\centering
\includegraphics[width=0.48\linewidth]{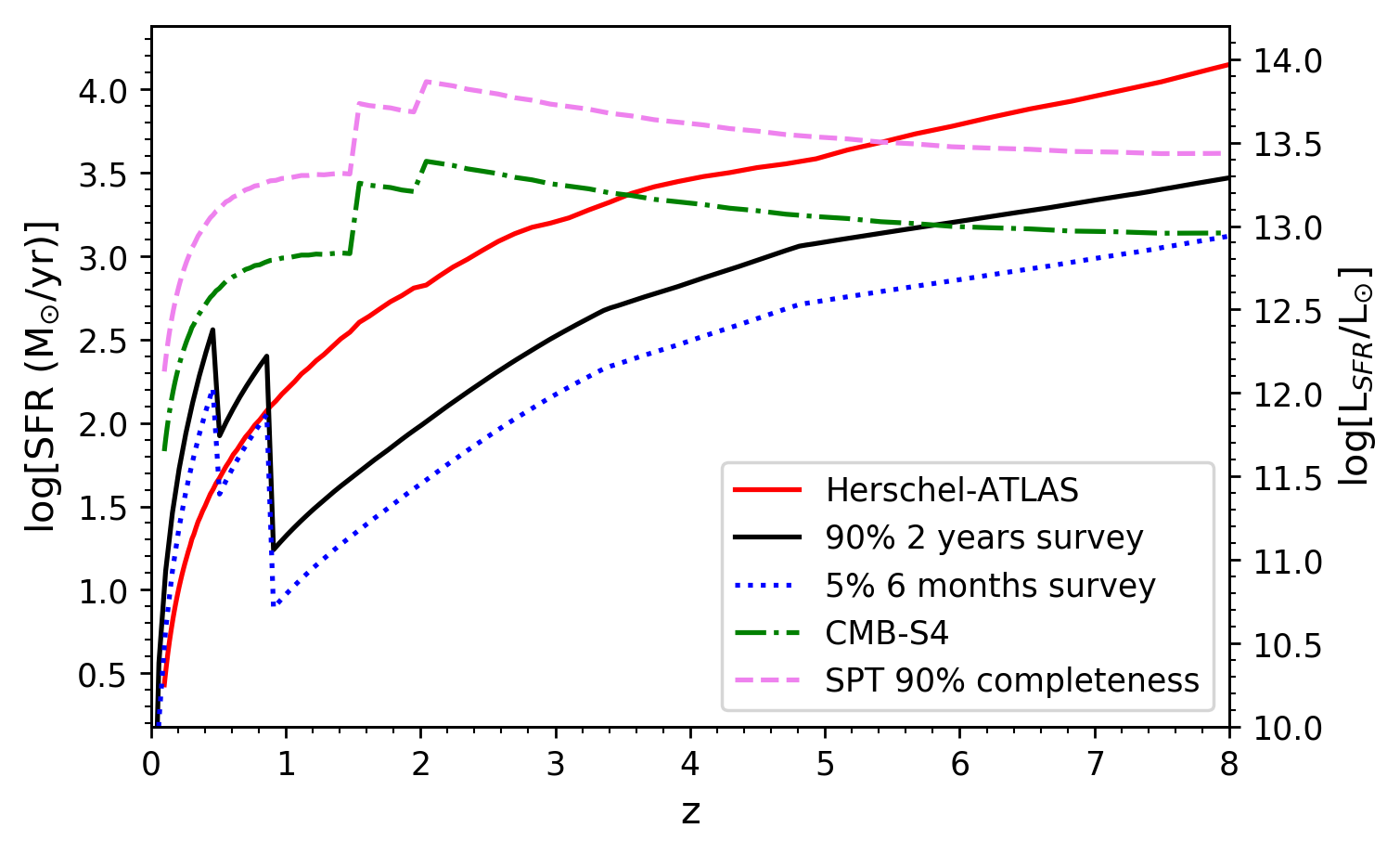}
\includegraphics[width=0.48\linewidth]{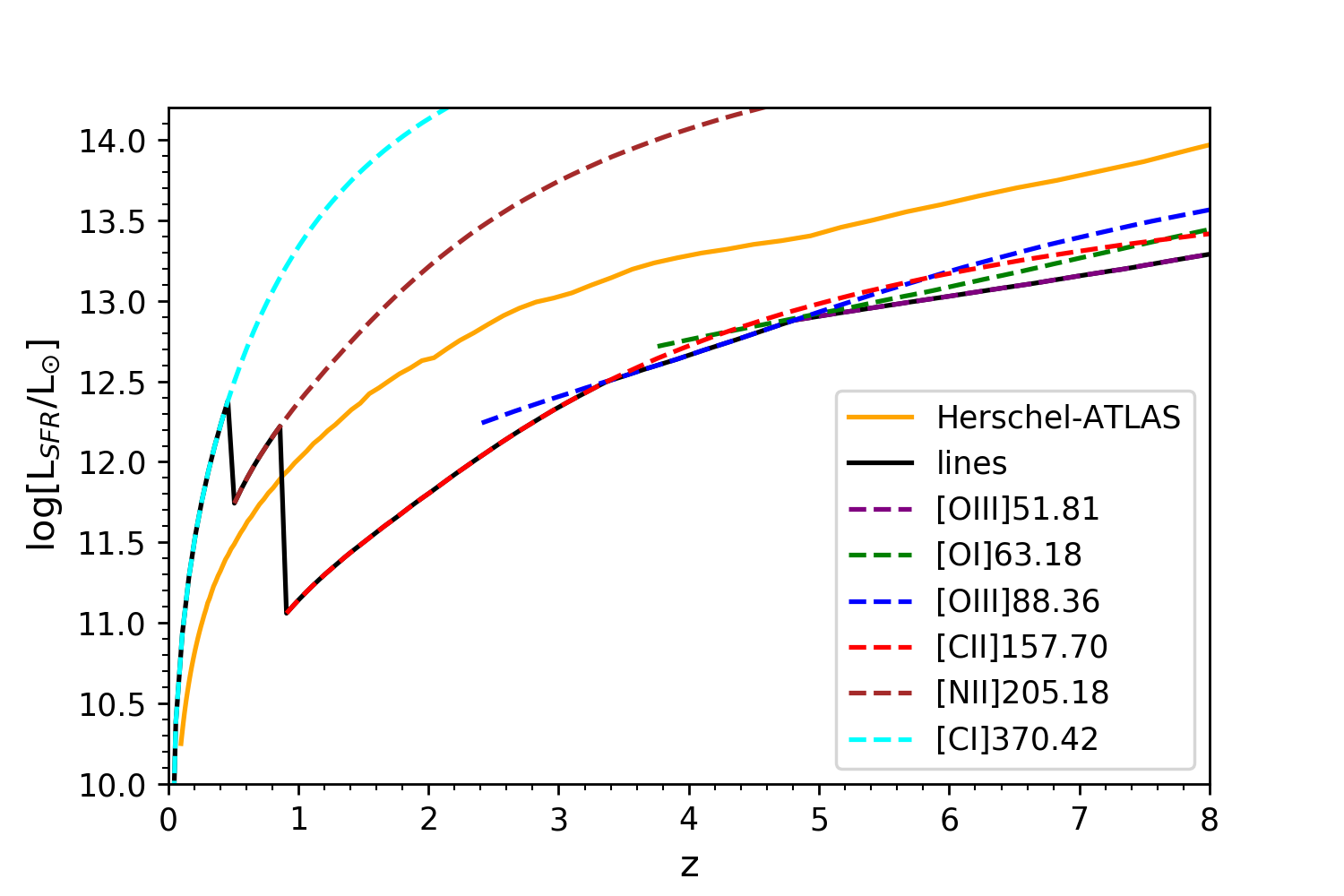}
\vspace{-4mm} \caption{\textit{Left}: Minimum SFR as a function of redshift,
for galaxies detected in lines in the 100--1000\,GHz range in an ``all-sky'' survey
(2 years, 90\,\% of the sky, solid black line) and for a deep survey  6-months duration
over 5\% of the sky (dotted blue line); see text for details. The scale
on the right refers to the bolometric luminosity due to star formation, $L_{\rm SFR}$,
based on the calibration by \citet{KennicuttEvans2012}.
The solid red line, the green dot-dashed line, and the magenta dashed line show, for comparison,
the IR (8--$1000\,\mu$m) luminosity, $L_{\rm IR}$,
corresponding to the $4\,\sigma$ detection limits (approximately 90\,\% completeness)
of the H-ATLAS survey covering $660\,\hbox{deg}^2$, to confusion limit of the CMB-S4 survey at 220\,GHz
expected to cover 43\% of the sky \citep[5\,mJy;][]{Abazajian2019}, and to the 90\,\% completeness limit (15\,mJy) of the
South Pole Telescope \citep[SPT;][]{Mocanu2013} survey covering $2,500\,\hbox{deg}^2$, respectively.
Here $L_{\rm IR}$ is a measure of the dust-obscured SFR.
\textit{Right}: Minimum $L_{\rm SFR}$ (or SFR) corresponding to the $5\,\sigma$ detection limits of the
brightest IR/submm lines over the 100--1000\,GHz range, for the ``all-sky'' survey.}
\label{fig:LIR_lim}
\end{figure*}

This instrument would deliver unique results in many branches of astrophysics.
Here we will specifically discuss its potential for:
\begin{itemize}
\item investigating the physical processes driving the assembly of galaxies
    and exploring the evolution of their metal and dust content out to
    $z\,{\simeq}\,8$;

\item measuring the early growth (to $z\,{\simeq}\,7$) of large-scale
    structures  (i.e., galaxy proto-clusters), when their member galaxies
    were actively star forming, and when the hot gas, making them detectable
    in X-rays or via the SZ effect, was not necessarily in place yet.
\end{itemize}
The plan of the paper is the following.  In Sect.~\ref{sect:det_lim} we give
the $5\,\sigma$ line-detection limits of the proposed instrument and the
corresponding minimum star-formation rate (SFR), as a function of redshift,
detectable in lines for the average relationship between line luminosity and
SFR. In Sect.~\ref{sect:SFRfunc} we briefly describe our reference model and
present predictions of the redshift distributions of galaxies detected in lines
and of the cumulative SFR functions at various redshifts, out to
$z\,{\simeq}\,8$. In Sects.~\ref{sect:lensed} and \ref{sect:unlensed} we
highlight examples of the new science enabled by these data on strongly
gravitationally lensed and unlensed galaxies. In Sect.~\ref{sect:comparison} we
compare the proposed survey with those of other forthcoming or planned
instruments. Section~\ref{sect:protocluster} discusses the potential of this
project for reconstructing the full history of the most massive virialized
structures in the Universe, namely galaxy clusters.
Section~\ref{sect:cluster_surveys} presents a comparison with cluster surveys
in other wavebands. Finally, Sect.~\ref{sect:conclusions} summarizes our main
conclusions.

We adopt a flat $\Lambda$CDM cosmology  with the latest values of the
parameters derived from Planck CMB power spectra: $H_0 =
67.4\,\hbox{km}\,\hbox{s}^{-1}\, \hbox{Mpc}^{-1}$; and $\Omega_{\rm m} = 0.315$
\citep{PlanckCollaboration2018parameters}.

\begin{table}
\caption{Sensitivity in temperature, $\Delta T$ (in units of temperature times square root of solid angle),
point source detection limits, $S_{\rm lim}$, and line-detection limits,
$\log(F_{\rm lim})$, at selected frequencies, $\nu$, (or wavelengths $\lambda$)
for a 2-yr survey of 90\,\% of the sky with the instrument described in the text.
Both $S_{\rm lim}$ and $F_{\rm lim}$ are at the $5\,\sigma$ significance level.
The angular resolution of the instrument, measured by the FWHM (full-width at half maximum
of the beam) at each frequency, is at the diffraction limit.
}
\label{tab:sensitivities}
\resizebox{0.49\textwidth}{!}{
\setlength{\tabcolsep}{4pt}
\begin{tabular}{rrrrrr}
\hline
\noalign{\vskip 3pt}
  \multicolumn{1}{c}{$\nu$} &
  \multicolumn{1}{c}{$\lambda$} &
  \multicolumn{1}{c}{FWHM} &
  \multicolumn{1}{c}{$\Delta T$} &
  \multicolumn{1}{c}{$S_{\rm lim}$} &
  \multicolumn{1}{c}{$\log(F_{\rm lim})$} \\
  \multicolumn{1}{c}{[GHz]} &
  \multicolumn{1}{c}{[mm]} &
  \multicolumn{1}{c}{[arcmin]} &
  \multicolumn{1}{c}{[$\mu\hbox{K}\cdot$arcmin]} &
  \multicolumn{1}{c}{[mJy]} &
  \multicolumn{1}{c}{[$\hbox{W}\,\hbox{m}^{-2}$]} \\
  \noalign{\vskip 3pt}
  \hline
  100	 & 	2.998	 & 	3.80	 & 	58.70	 & 	30.87	 & 	-18.988	 \\
120	 & 	2.498	 & 	3.17	 & 	51.86	 & 	32.73	 & 	-18.883	 \\
140	 & 	2.141	 & 	2.72	 & 	45.76	 & 	33.69	 & 	-18.803	 \\
160	 & 	1.874	 & 	2.38	 & 	40.38	 & 	33.98	 & 	-18.742	 \\
180	 & 	1.666	 & 	2.11	 & 	35.51	 & 	33.61	 & 	-18.695	 \\
200	 & 	1.499	 & 	1.90	 & 	31.18	 & 	32.80	 & 	-18.660	 \\
220	 & 	1.363	 & 	1.73	 & 	27.33	 & 	31.62	 & 	-18.635	 \\
240	 & 	1.249	 & 	1.58	 & 	23.91	 & 	30.18	 & 	-18.617	 \\
260	 & 	1.153	 & 	1.46	 & 	20.88	 & 	28.55	 & 	-18.607	 \\
280	 & 	1.071	 & 	1.36	 & 	18.23	 & 	26.84	 & 	-18.601	 \\
300	 & 	0.999	 & 	1.27	 & 	15.89	 & 	25.07	 & 	-18.601	 \\
320	 & 	0.937	 & 	1.19	 & 	13.81	 & 	23.24	 & 	-18.606	 \\
340	 & 	0.882	 & 	1.12	 & 	12.02	 & 	21.49	 & 	-18.613	 \\
360	 & 	0.833	 & 	1.06	 & 	10.47	 & 	19.82	 & 	-18.624	 \\
380	 & 	0.789	 & 	1.00	 & 	9.10	 & 	18.18	 & 	-18.638	 \\
400	 & 	0.749	 & 	0.95	 & 	7.94	 & 	16.71	 & 	-18.652	 \\
420	 & 	0.714	 & 	0.91	 & 	6.91	 & 	15.27	 & 	-18.670	 \\
440	 & 	0.681	 & 	0.86	 & 	6.05	 & 	14.01	 & 	-18.687	 \\
460	 & 	0.652	 & 	0.83	 & 	5.30	 & 	12.81	 & 	-18.707	 \\
480	 & 	0.625	 & 	0.79	 & 	4.67	 & 	11.78	 & 	-18.725	 \\
500	 & 	0.600	 & 	0.76	 & 	4.11	 & 	10.82	 & 	-18.744	 \\
520	 & 	0.577	 & 	0.73	 & 	3.66	 & 	10.00	 & 	-18.761	 \\
540	 & 	0.555	 & 	0.70	 & 	3.26	 & 	9.27	 & 	-18.778	 \\
560	 & 	0.535	 & 	0.68	 & 	2.93	 & 	8.63	 & 	-18.793	 \\
580	 & 	0.517	 & 	0.66	 & 	2.65	 & 	8.09	 & 	-18.806	 \\
600	 & 	0.500	 & 	0.63	 & 	2.42	 & 	7.64	 & 	-18.816	 \\
620	 & 	0.484	 & 	0.61	 & 	2.22	 & 	7.24	 & 	-18.825	 \\
640	 & 	0.468	 & 	0.59	 & 	2.06	 & 	6.92	 & 	-18.831	 \\
660	 & 	0.454	 & 	0.58	 & 	1.92	 & 	6.67	 & 	-18.833	 \\
680	 & 	0.441	 & 	0.56	 & 	1.80	 & 	6.43	 & 	-18.836	 \\
700	 & 	0.428	 & 	0.54	 & 	1.70	 & 	6.25	 & 	-18.836	 \\
720	 & 	0.416	 & 	0.53	 & 	1.62	 & 	6.12	 & 	-18.833	 \\
740	 & 	0.405	 & 	0.51	 & 	1.54	 & 	6.00	 & 	-18.830	 \\
760	 & 	0.394	 & 	0.50	 & 	1.47	 & 	5.88	 & 	-18.827	 \\
780	 & 	0.384	 & 	0.49	 & 	1.42	 & 	5.83	 & 	-18.819	 \\
800	 & 	0.375	 & 	0.48	 & 	1.37	 & 	5.77	 & 	-18.813	 \\
820	 & 	0.366	 & 	0.46	 & 	1.33	 & 	5.72	 & 	-18.806	 \\
840	 & 	0.357	 & 	0.45	 & 	1.29	 & 	5.68	 & 	-18.798	 \\
860	 & 	0.349	 & 	0.44	 & 	1.25	 & 	5.66	 & 	-18.790	 \\
880	 & 	0.341	 & 	0.43	 & 	1.22	 & 	5.65	 & 	-18.781	 \\
900	 & 	0.333	 & 	0.42	 & 	1.19	 & 	5.63	 & 	-18.772	 \\
920	 & 	0.326	 & 	0.41	 & 	1.16	 & 	5.63	 & 	-18.763	 \\
940	 & 	0.319	 & 	0.40	 & 	1.14	 & 	5.62	 & 	-18.754	 \\
960	 & 	0.312	 & 	0.40	 & 	1.11	 & 	5.63	 & 	-18.744	 \\
980	 & 	0.306	 & 	0.39	 & 	1.09	 & 	5.63	 & 	-18.735	 \\
1000	 & 	0.300	 & 	0.38	 & 	1.07	 & 	5.64	 & 	-18.726	 \\
  \noalign{\vskip 3pt}
  \hline\end{tabular}     }
  \end{table}

\section{Line-detection limits}\label{sect:det_lim}

Table~\ref{tab:sensitivities} shows  the estimated sensitivity and the
$5\,\sigma$ point source and line-detection limits at a set of frequencies for
a 2-year survey of 90\,\% of the sky with 64 polarized channelizers, covering
the frequency range 100--1000\,GHz with $R\,{=}\,300$, using close to
background limited MKID detectors (we assume a conservative 30\,\% optical
efficiency, and instrumental noise matching the sky background noise). A
6-month duration survey of 5\,\% sky is also being considered; it goes deeper
by a factor of approximately $\sqrt{5}$.

\begin{figure*}[htbp!]
\vspace{-4mm}
\centering
\includegraphics[width=0.49\linewidth]{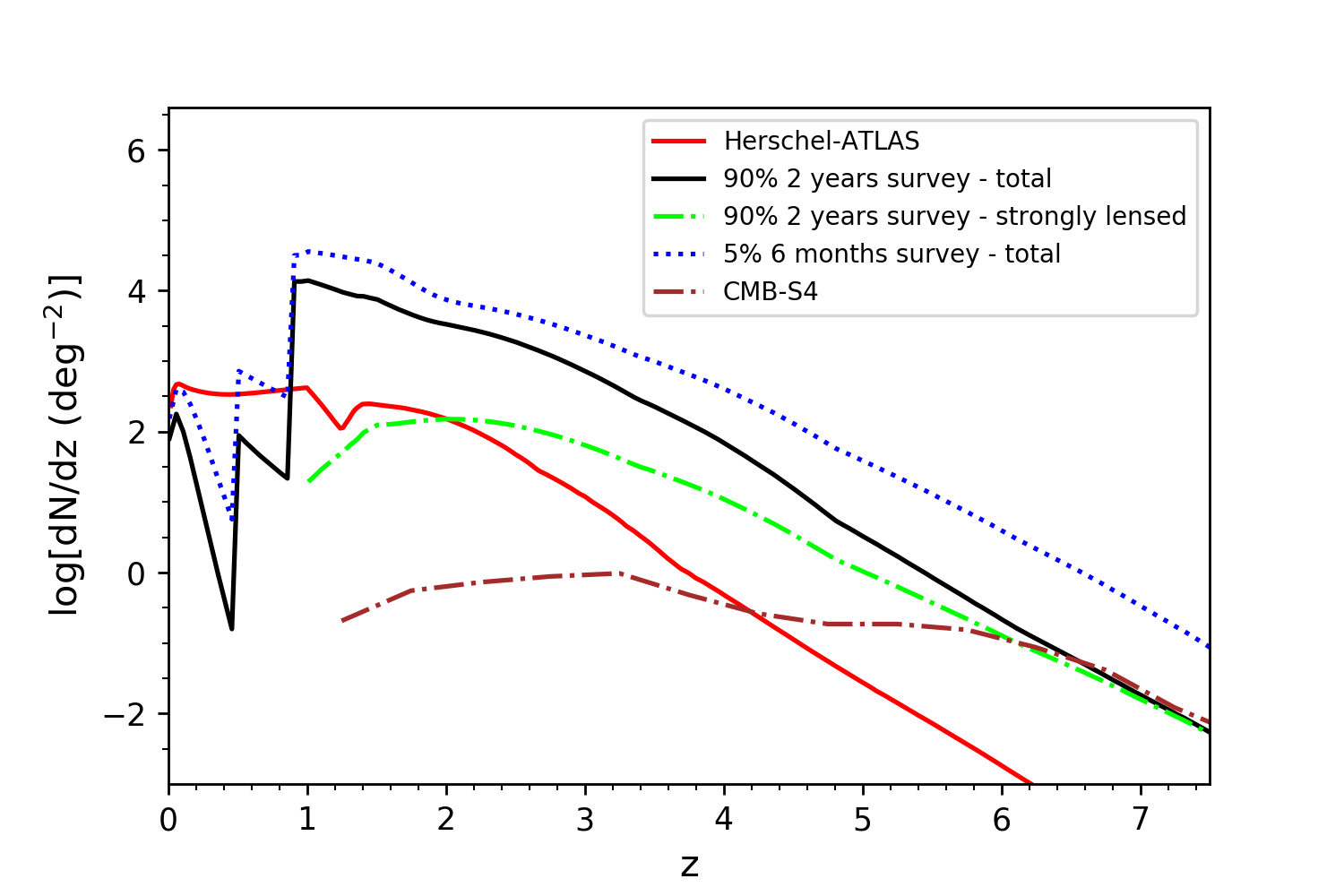}
\includegraphics[width=0.49\linewidth]{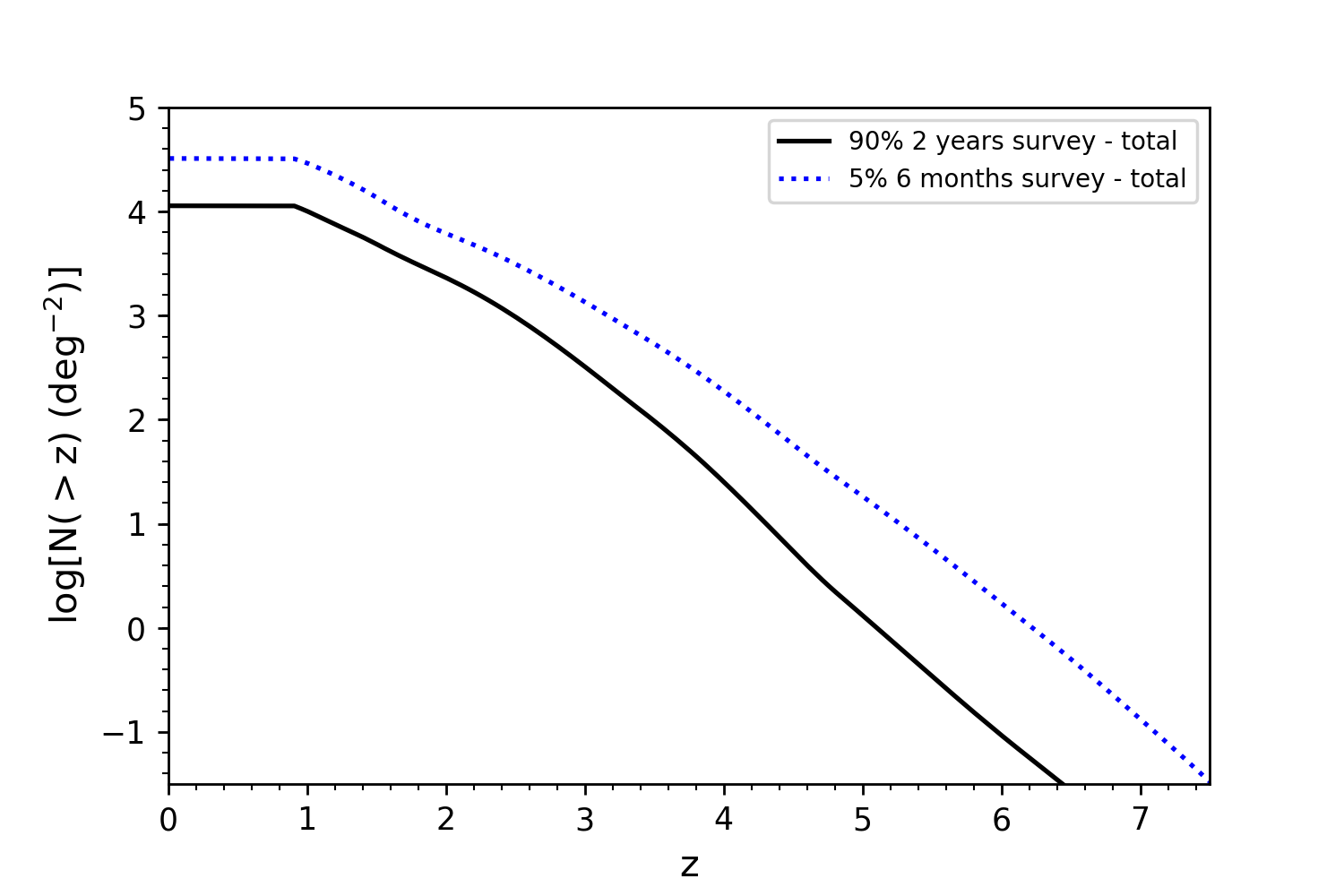}
\caption{\textit{Left}: Predicted differential redshift distributions of galaxies detected in
at least one line by the ``all-sky'' survey (2\,yr, 90\,\% of the sky, with the solid black
line being the total and the dot-dashed green line being for strongly lensed galaxies) and
by the ``deep'' survey (6 months, 5\,\% of the sky, with the dotted blue line
showing the total distribution).  For comparison the solid red line
shows the estimated redshift distribution of galaxies
detected by the H-ATLAS survey over $660\,\hbox{deg}^2$ above the $4\,\sigma$ limit in at least one
SPIRE channel, based on the \citet{Cai2013} model. The dot-dashed brown line
shows the predicted redshift distribution at the confusion limit of the CMB-S4
survey (with an expected sky coverage of 43\%), derived from the cumulative distribution in figure~26 of the CMB-S4 Science Case paper
\citep{Abazajian2019}. \textit{Right}: Total cumulative redshift distributions for the ``all-sky''
and for the ``deep'' survey (solid black line and dotted blue lines, respectively.} \label{fig:gal_zdistr}
\end{figure*}

Exploiting observations, mostly from \textit{Spitzer} and \textit{Herschel},
\citet{Bonato2019} reported tight correlations between the main mid-IR-to-submm
lines from neutral or ionized atomic gas and from molecular gas and the total
IR luminosity, $L_{\rm IR}$ (conventionally defined over 8--$1000\,\mu$m), of
dust heated by newly formed stars.  The brightest lines detectable in the
frequency range considered here are [{\sc Nii}]\,205.18\,$\mu$m,  [{\sc
Ci}]\,170.42\,$\mu$m, [{\sc Cii}]\,157.7\,$\mu$m, [{\sc Oiii}]\,88.36\,$\mu$m,
[{\sc Oi}]\,63.18\,$\mu$m and [{\sc Oiii}]\,51.81\,$\mu$m.

The line--$L_{\rm IR}$ relations obtained by \citet{Bonato2019} are based  on
observations of dusty galaxies, for which the unabsorbed fraction of the
emission from young stars is small, so that $L_{\rm IR}$ is a measure of the
total SFR. One might wonder whether the line luminosities are primarily related
to $L_{\rm IR}$ or to the SFR and an answer was provided by
\citet{DeLooze2014}. For their sample of low-metallicity galaxies with moderate
dust emission they found that [{\sc Oiii}]\,88.36\,$\mu$m and [{\sc
Oi}]\,63.18\,$\mu$m are good probes of the SFR measured through a composite
tracer ({\it GALEX\/} FUV + {\it Spitzer}-MIPS\,$24\,\mu$m luminosity). The
case is less clear for [{\sc Cii}]\,157.7\,$\mu$m; however, the [{\sc Cii}]
emission in star-forming galaxies primarily arises from photo-dissociation
regions \citep[PDRs;][]{Stacey2010}, although it can also come from the partly
ionized interstellar medium \citep[e.g.,][]{Sutter2019}. Since PDRs are heated
by the UV radiation emitted by young stars, [{\sc Cii}] has also been used as a
measure of the SFR \citep[e.g.,][]{Carniani2018}. The [{\sc Cii}] luminosities
of the $z\,{\simeq}\,6.8$ galaxies without detected FIR emission, observed by
\citet{Smit2018} were found to be good SFR estimators. \citet{Schaerer2020}
reported no or weak evolution of the [{\sc Cii}]--SFR relation over the last
13\,Gyr, i.e. up to $z\sim 8$.

Based on these results, following \citet{Bonato2019} we assume that the
luminosity  of FIR fine-structure lines primarily correlates with the SFR.
Thus, the relations derived by \citet{Bonato2019} allow us to estimate the
values of SFR as a function of $z$ corresponding to the $5\,\sigma$
line-detection limits listed in Table~\ref{tab:sensitivities}. The results are
shown in Fig.~\ref{fig:LIR_lim}, where the solid black line refers to the
survey of 90\,\% of the sky (the ``all-sky'' survey), while the dotted blue
line refers to the deep survey of 5\,\% of the sky. At each redshift we have
computed the minimum SFR detectable in the various lines and taken the smallest
one.

Our calculations assume that the lines are unresolved at $R=300$, corresponding
to line widths of $1,000\,\hbox{km}\,\hbox{s}^{-1}$. This is almost always the
case for individual galaxies. The spectroscopy of 15 H-ATLAS galaxies at $2.08
< z < 4.05$ obtained by \citet{Neri2019} with the IRAM NOrthern Extended
Millimeter Array (NOEMA) yielded line widths (FWHM) between 150 and
$1100\,\hbox{km}\,\hbox{s}^{-1}$ (mean $700\pm 300\,\hbox{km}\,\hbox{s}^{-1}$,
median $800\,\hbox{km}\,\hbox{s}^{-1}$).  The [{\sc Cii}]\,157.7\,$\mu$m
observations by \citet{Gullberg2015} of 16 spectrally resolved strongly lensed
star-forming galaxies at $3.0 < z < 5.7$ yielded line FWHMs ranging from
$198\pm 34$ to $800\pm 200\,\hbox{km}\,\hbox{s}^{-1}$ with a median of $\simeq
541\pm 110\,\hbox{km}\,\hbox{s}^{-1}$. \citet{Nesvadba2019} measured the FWHMs
of the [{\sc Oi}] $370.42\,\mu$m and $609.14\,\mu$m lines in the range from
$220\pm 21$ to $639\pm 100\,\hbox{km}\,\hbox{s}^{-1}$  (median
$475\,\hbox{km}\,\hbox{s}^{-1}$) for 11 strongly lensed submm galaxies detected
by \textit{Planck}. Similar values (from $220\pm 50$ to $770\pm
80\,\hbox{km}\,\hbox{s}^{-1}$, median $370\,\hbox{km}\,\hbox{s}^{-1}$) were
measured by \citet{Cooke2018} for 10 serendipitous [{\sc Cii}]\,157.7\,$\mu$m
emitters at $z\sim 4.5$.

A somewhat lower median FWHM ($252\,\hbox{km}\,\hbox{s}^{-1}$) was reported by
\citet{Bethermin2020} for 75 ALPINE-ALMA large program targets at $4.4 < z <
5.9$, detected in the [{\sc Cii}]\,157.7\,$\mu$m line. These targets have
substantially lower SFRs than the strongly lensed galaxies observed by
\citet{Gullberg2015} and \citet{Nesvadba2019}. A result close to that by
\citet{Bethermin2020} was found by \citet{Fujimoto2019} for a sample of 18
galaxies at a higher mean redshift ($5.153\le z \le 7.142$): the weighted
average FWHM of the [{\sc Cii}]\,157.7\,$\mu$m line for their sample is
$270\,\hbox{km}\,\hbox{s}^{-1}$.

At still higher redshifts, \citet{Tamura2019} and \citet{Hashimoto2018}
reported lower FWHMs of the [{\sc Oiii}]\,88.36\,$\mu$m line for galaxies at
$z=8.31$ ($141\pm 21\,\hbox{km}\,\hbox{s}^{-1}$) and at $z=9.11$ ($154\pm
39\,\hbox{km}\,\hbox{s}^{-1}$), respectively. An even lower line width ($\simeq
43\,\hbox{km}\,\hbox{s}^{-1}$) was derived by \citet{Laporte2017} for a galaxy
at $z=8.38$.

Thus, for ultraluminous submm galaxies the signal dilution due to the modest
spectral resolution of the instrument is moderate, up to $z\simeq 5$--6, while
it is stronger at higher $z$ and at lower luminosities.

The information on the velocity dispersion of galaxies in high-$z$
proto-cluster \textit{cores} of submm galaxies is still very limited.
\citet{Hill2020} estimated a line-of-sight velocity dispersion $\sigma_r=376\pm
68\,\hbox{km}\,\hbox{s}^{-1}$ for the $z=4.3$ proto-cluster core discovered by
\citet{Miller2018}. \citet{Oteo2018} found a substantially higher $\sigma_r$
($794\pm 68\,\hbox{km}\,\hbox{s}^{-1}$) for their $z=4.0$ proto-cluster core.
However, they argued that the detected galaxies actually belong to two groups,
each with a much lower $\sigma_r$. \citet{Venemans2007} measured velocity
dispersions of forming clusters of galaxies near powerful radio galaxies at
$2.0 < z < 5.2$. They found that $\sigma_r$ increases from $\sim
300\,\hbox{km}\,\hbox{s}^{-1}$ at $z>4$ to
$500$--$700\,\hbox{km}\,\hbox{s}^{-1}$ at $z \sim 3$. In the two lowest
redshift fields ($z=2.86$ and $z=2.16$) the velocity distribution is bimodal,
indicating the presence of subgroups with $\sigma_r
=200$--$500\,\hbox{km}\,\hbox{s}^{-1}$. This shows that, on one side, the lines
from the spatially unresolved cluster cores are spectrally unresolved except at
the lowest redshifts and, on the other hand, that the signal dilution is
expected to be from moderate to low.

The results for the brightest lines are shown in the right-hand panel of
Fig.~\ref{fig:LIR_lim}. The best lines are [{\sc Ci}]\,170.42\,$\mu$m for
$z\,{\simlt}\,0.5$, [{\sc Nii}]\,205.18\,$\mu$m for
$0.5\,{\simlt}\,z\,{\simlt}\,0.9$, [{\sc Cii}]\,157.7\,$\mu$m for
$0.9\,{\simlt}\,z\,{\simlt}\,3.4$, [{\sc Oiii}]\,88.36\,$\mu$m for
$3.4\,{\simlt}\,z\,{\simlt}\,4.8$ and [{\sc Oiii}]\,51.81\,$\mu$m for
$z\,{\simgt}\,4.8$.

The right-hand panel of Fig.~\ref{fig:LIR_lim} shows that the detection limits
for  [{\sc Cii}]\,157.7\,$\mu$m, [{\sc Oiii}]\,88.36\,$\mu$m, [{\sc
Oi]}\,63.18\,$\mu$m, and [{\sc Oiii}]\,51.81\,$\mu$m correspond to values of
SFR that are quite close to each other. Hence a substantial fraction of
galaxies will be detected both in the [{\sc Cii}]\,157.7\,$\mu$m and [{\sc
Oiii}]\,88.36\,$\mu$m lines for $z\,{\simgt}\,2.4$ (when the second line shows
up at $\nu\,{<}\,1000\,$GHz). The [{\sc Oi}]\,63.18\,$\mu$m and [{\sc
Oiii}]\,51.81\,$\mu$m lines come in at $z\,{\ge}\,3.7$ and $z\,{\ge}\,4.8$,
respectively.

For comparison, the solid red line in the left-hand panel of
Fig.~\ref{fig:LIR_lim}  also shows the minimum $L_{\rm IR}$, or the minimum
dust-enshrouded SFR, among those corresponding to the $4\,\sigma$ detection
limits of the H-ATLAS survey  \citep[the largest extragalactic survey with
\textit{Herschel}, having sensitivities of 29.4, 37.4, and 40.6\,mJy at 250,
350, and $500\,\mu$m, respectively;][]{Valiante2016}. The monochromatic
luminosities corresponding to these detection limits have been converted to
$L_{\rm IR}$ using the spectral energy distributions (SEDs) adopted by
\citet{Cai2013}. More precisely, for $z\,{<}\,1.5$ we have used the ``warm''
(starburst) SED and at $z\,{>}\,2$ the ``proto-spheroidal'' SED; at
intermediate redshifts we considered both SEDs and chose the more favourable
one, i.e., the one yielding the lower $L_{\rm IR}$. Since the submm continuum
measures only the light re-emitted by dust, while the fine-structure lines
measure the total SFR, the comparison of the two measurements, both made by the
proposed instrument, provides information on the effective optical depths of
galaxies.

\begin{figure*}[htbp!]
\includegraphics[width=\textwidth]{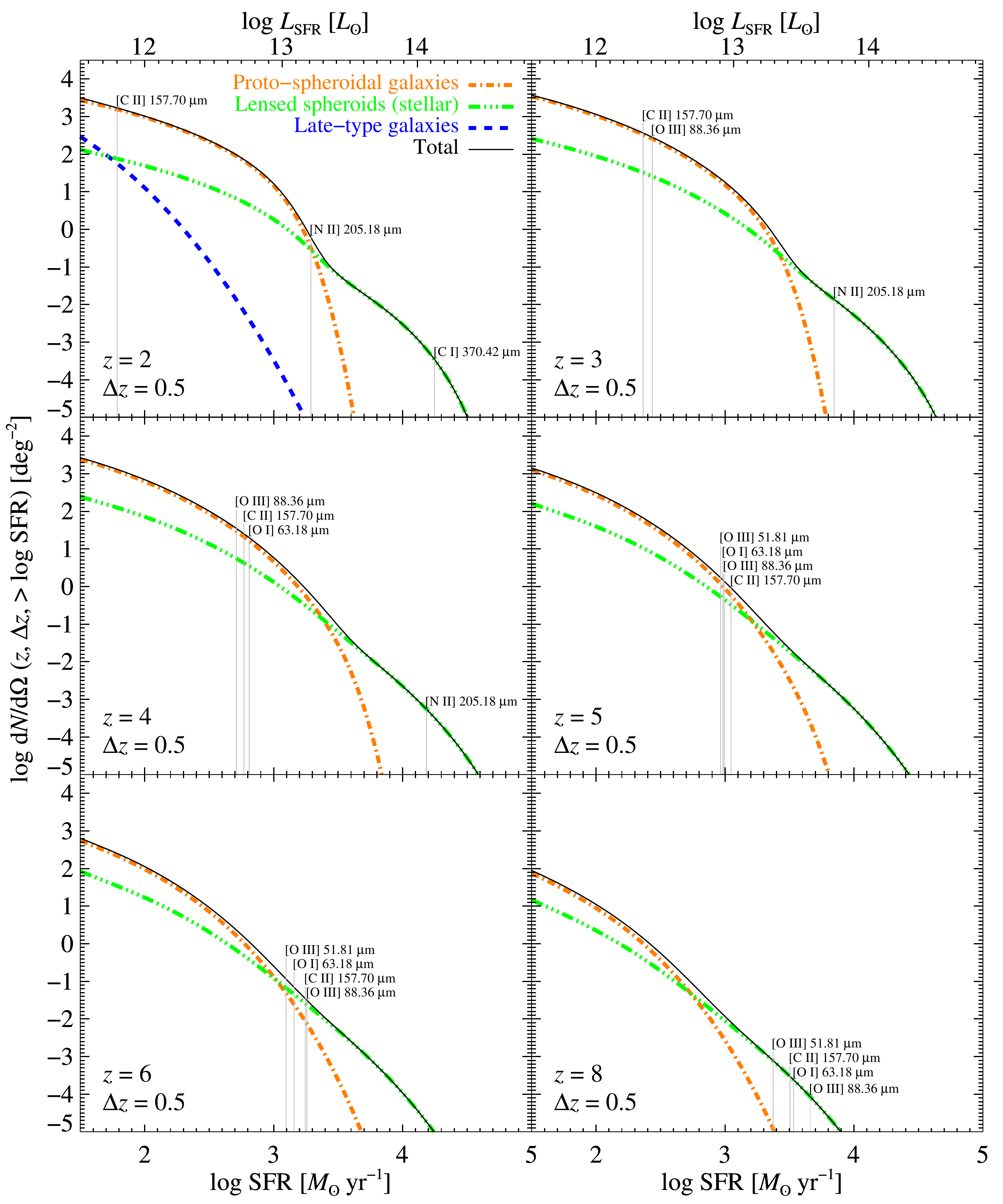}
\caption{Cumulative SFR functions of galaxies
within $\delta z= 0.5$ derived from the \citet{Cai2013} model at redshifts from 2 to 8. The
vertical lines show the SFRs corresponding to the $5\,\sigma$
line-detection limits of the brightest lines. }
\label{fig:galLFs}
\end{figure*}

\section{SFR functions and redshift distributions}\label{sect:SFRfunc}

To investigate quantitatively the discovery potential of the proposed
spectroscopic survey, the relationships between line luminosity and SFR need to
be coupled with a model for the redshift-dependent SFR function. We adopted the
physically grounded model by \citet{Cai2013}. This model is built on the
consideration that, in the local Universe, spheroids (i.e., ellipticals and
bulges of disk galaxies) are mostly inhabited by old stellar populations
(formation redshifts $z\gtrsim 1$--1.5), while the populations of disk galaxies
are generally younger, with luminosity-weighted ages mostly $\simlt 7\,$Gyr
\citep[cf., e.g.,][]{Bernardi2010}, corresponding to formation redshifts
$z\simlt 1$. Thus, spheroid progenitors (referred to as proto-spheroids or
proto-spheroidal galaxies) are the dominant star-forming population at
$z\gtrsim 1.5$, whereas most of the star formation at $z\simlt 1.5$ occurs in
disks.

These different evolutionary histories are dealt with adopting a ``hybrid''
approach. The model provides a physically grounded description of the
redshift-dependent co-evolution of the SFR of proto-spheroidal galaxies and of
the the active nuclei at their centers, while the description of the evolution
of late-type galaxies and of AGN associated with them is phenomenological and
parametric.

The calculation of the evolving SFR function of proto-spheroids hinges upon the
halo formation rate as a function of redshift, $z$, and of halo mass, $M_{\rm
H}$, provided by large-scale $N$-body simulations. The analytical approximation
of the halo mass function, $N(M_{\rm H},z)$, derived by \citet{ShethTormen1999}
was used. The positive term of its time derivative was adopted as a good
approximation of the halo formation rate. High resolution $N$-body simulations
\citep[e.g.,][]{Wang2011} showed that, after a fast collapse phase, including
major mergers, the halo growth (mostly by minor mergers and diffuse accretion)
mainly affects the halo outskirts and has little impact on the inner potential
well where the visible galaxy resides. Based on these results, the model
assumes that the main drivers of star formation and AGN growth are in-situ
processes. The star-formation history of proto-spheroids is computed by solving
a set of equations describing the evolution of gas phases and of the active
nucleus, including cooling, condensation into stars, radiation drag, accretion,
and feedback from supernovae and from the AGN.

Solving these equations, we obtain the SFR of each galaxy and the bolometric
luminosity of the AGN as a function of halo mass, formation redshift, and
galactic age. Coupling the $\hbox{SFR}(M_{\rm H},z)$ with the halo formation
rate we get the SFR function of proto-spheroids at any redshift. As for
late-type galaxies, \citet{Cai2013} adopted a phenomenological evolutionary
model, with different parameters for starburst and ``normal'' disk galaxies.
The global (proto-spheroids plus late-type galaxies) SFR functions yielded by
the model are in excellent agreement with observational determinations obtained
combining information from far-IR/submm, UV and Ly$\,\alpha$ surveys
\citep{Cai2014, Mancuso2015}. Therefore, the results based on them are
essentially model independent.

Figures~\ref{fig:gal_zdistr} and \ref{fig:galLFs} show that the spectroscopic
survey will allow us to extend the study of the global star-formation history
and the build-up of metals and dust all the way through the epoch of
reionization. The survey will detect thousands of star-forming galaxies at
$z\,{\simeq}\,6$ and several tens at $z\,{\simeq}\,8$. Note that the predicted
abundances of high-$z$ galaxies reported in these figures may be underestimated
if the stellar initial mass function (IMF) becomes more top-heavy (i.e., has a
larger fraction of massive stars compared to standard IMFs) at high-$z$, as
suggested by theoretical arguments \citep[e.g.,][]{Papadopoulos2011} and
indicated by some observational evidence \citep{Zhang2018}. A more top-heavy
IMF would yield higher surface densities of ultra-luminous high-$z$ galaxies.
An excess of $z\,{\simgt}\,4$ dusty galaxies over model expectations has been
reported, but the issue is controversial \citep[see][for a recent discussion
and references]{Cai2019}.

The excellent sensitivity of the instrument, the high luminosity of the [{\sc
Cii}]\,157.7\,$\mu$m, [{\sc Oiii}]\,88.36\,$\mu$m, [{\sc Oi}]\,63.18\,$\mu$m,
and [{\sc Oiii}]\,51.81\,$\mu$m lines and the immunity to the confusion limit
make the spectroscopic survey far more efficient than broadband surveys at
detecting high-$z$ star-forming galaxies.  This is illustrated by the
comparison with the estimated redshift distribution of H-ATLAS galaxies.

The redshift distribution of the spectroscopic survey
(Fig.~\ref{fig:gal_zdistr})  peaks at $z\,{\simeq}\,1$--1.5. The brightest line
within the frequency range covered by the instrument varies with redshift (see
the right-hand panel of Fig.~\ref{fig:LIR_lim}). These variations produce the
indentations at low $z$ and the secondary peak at $z\simeq 0.6$. The solid red
line shows, for comparison, the estimated redshift distribution of galaxies
detected by the H-ATLAS survey above the $4\,\sigma$ limit in at least one
SPIRE channel (based on the \citealt{Cai2013} model). The dip at
$z\,{\simeq}\,1.5$ corresponds to the change of the dominant star-forming
population: late-type plus starburst galaxies and proto-spheroidal galaxies at
lower and higher redshifts, respectively.

\section{Strongly lensed galaxies}\label{sect:lensed}

Figures~\ref{fig:gal_zdistr} and \ref{fig:galLFs} also show that most
spectroscopically detected $z\,{\simgt}\,4$ galaxies and the brightest galaxies
at lower redshifts are strongly lensed \citep[see discussions
in][]{Perotta2002,Negrello2007,Paciga2009,Lima2010}. The availability of large
samples of strongly lensed galaxies out to high redshifts will drive a real
breakthrough in the study of the early evolutionary phases of galaxies. The
{\it Herschel} surveys H-ATLAS and HerMES, the {\it Planck} all-sky survey and
the SPT survey have already provides several hundreds of lensed galaxy
candidates, with painstaking follow-up campaigns world-wide and at all
wavelengths \citep[e.g.,][]{Negrello2010, Negrello2014, Negrello2017lensed,
Lupu2012, Vieira2013, Wardlow2013, Hezaveh2013, Canameras2015, Harrington2016,
Nayyeri2016, Harrington2018, Yang2017, DiazSanchez2017, Massardi2018, Bakx2018,
Dannerbauer2019a}.

Strong lensing not only boosts the observed global flux by a factor $\mu$ but
also increases the angular sizes of galaxies by an average factor of
$\mu^{1/2}$. Since the magnification $\mu$ can be several tens, the expansion
of the image can be quite substantial. The study in great detail of the
internal structure and kinematics of galaxies will then become accessible to
high resolution instruments like ALMA or the {\it James Webb Space Telescope\/}
({\it JWST\/}). 
A spectacular demonstration of the power of strong gravitational lensing in
this respect was provided by ALMA 0.1\,arcsec resolution observations of the
\textit{Planck} source PLCK\,G244.8$+$54.9 at $z\,{\simeq}\,3.0$, with
$\mu\,{\simeq}\,30$ \citep{Canameras2017ALMA}. These observations reached the
astounding spatial resolution of 60\,pc, comparable to the size of GMCs (around
40--100\,pc). Very recently, ALMA high resolution observations of a strongly
lensed, normal star-forming galaxy at $z\,{=}\,1.06$ (the ``Cosmic Snake'')
even reached a spatial resolution of 30\,pc in the source plane
\citep{Dessauges2019}. Intriguingly, the 17 identified GMCs in this source have
different physical properties on average than those of nearby galaxies.

AGN-driven outflows are a key ingredient of current galaxy evolution models
\citep[see e.g.,][]{Heckman2014}, since they provide the most plausible
explanation for the deviation of the galaxy stellar mass function from the halo
mass function at large masses, i.e., for the low star-formation efficiency in
massive halos, only less than 10\,\% of baryons initially present in such halos
are used to form stars. However, information on the effect of feedback on the
direct fuel for star formation (namely, molecular gas) during the epoch of the
most active cosmic star formation is largely missing, due to the weakness of
spectral signatures of molecular outflows.

Gravitational lensing allows us to overcome these difficulties.
\citet{Canameras2017ALMA} obtained CO spectroscopy with a velocity resolution
of 40--50\,${\rm km}\,{\rm s}^{-1}$. This spectral resolution makes possible a
direct investigation of massive outflows driven by AGN feedback at high $z$,
with predicted velocities of order $1000\,\hbox{km}\,\hbox{s}^{-1}$
\citep{KingPounds2015}. \citet{Spilker2018} and \citet{Jones2019} detected, by
means of ALMA spectroscopy, massive molecular outflows in two strongly lensed
galaxies at $z\,{=}\,5.3$ and $z\,{=}\,5.7$, respectively, discovered by SPT
survey. \citet{Canameras2018outflow} detected a molecular wind signature in the
strongly lensed galaxy PLCK\,G165.7+49.0, discovered by \textit{Planck},  at
$z\,{=}\,2.2$, with magnification factors between 20 and 50 over most of the
source of emission.


\section{Unlensed galaxy populations}\label{sect:unlensed}

The evolution of the dust-obscured star formation is still poorly known.
\citet{Gruppioni2013}, \citet{Magnelli2013} and \citet{WangL2019} estimated the
IR luminosity functions of galaxies based on the \textit{Herschel} PACS and
SPIRE survey data. Above $z\,{\simeq}\,2.5$ the overwhelming majority of
redshifts are photometric (at $z\,{>}\,3$ there are only about 4--6\,\%
spectroscopic redshifts, depending on the survey field). Substantially
increasing the fraction of spectroscopic redshifts is hard
because of the faintness of these galaxies in the optical/near-IR. 



All these studies required the use of optical/near-IR data for photometric
redshift estimates. This means that heavily dust-obscured galaxies
are missed. 
Recent investigations \citep{Wang2019,
Dudzeviciute2019, Williams2019} have shown that a substantial fraction of
massive galaxies at $z\,{>}\,3$ are optically dark.

The proposed spectroscopic survey will provide an unbiased determination of the
redshift-dependent IR luminosity function up to $z\,{\simeq}\,6$. For the first
time we will have spectroscopic redshifts for a huge number of lensed and
non-lensed DSFGs, obtained in an unbiased way, independent of the
identification/pre-selection method. Out to $z\,{\simeq}\,4$ the luminosity
functions will be determined down to below the characteristic $L_{\star,\rm
IR}$. At $z\,{\simgt}\,2.4$ we expect detections in both the [{\sc
Cii}]\,157.7\,$\mu$m and the [{\sc Oiii}]\,88.36\,$\mu$m line. 

Such an all sky survey offers the opportunity to search for non-lensed,
hyper-lumninous infrared galaxied (HyLIRGs; $L_{\rm IR}>10^{13} {\rm
L}_\odot$). Currently, only a few examples are known in the distant Universe
\citep{Fu2012, Ivison2013, Riechers2013}, whilst models predict a factor of 2
higher numbers of HyLIRGs \citep{Cai2013}. Furthermore, the fraction of HyLIRGs
within the brightest {\it Herschel} galaxies is still unknown.

We note that, in particular, far-IR fine-structure lines seem to be the best
way to obtain spectroscopic redshifts of rest-frame UV-selected galaxies beyond
$z\,{=}\,7$ \citep[e.g.,][]{Inoue2016, Carniani2017, Hashimoto2018,
Laporte2017, Smit2018, Tamura2019}.

\begin{figure}[htbp!]
\centering
\includegraphics[width=\linewidth]{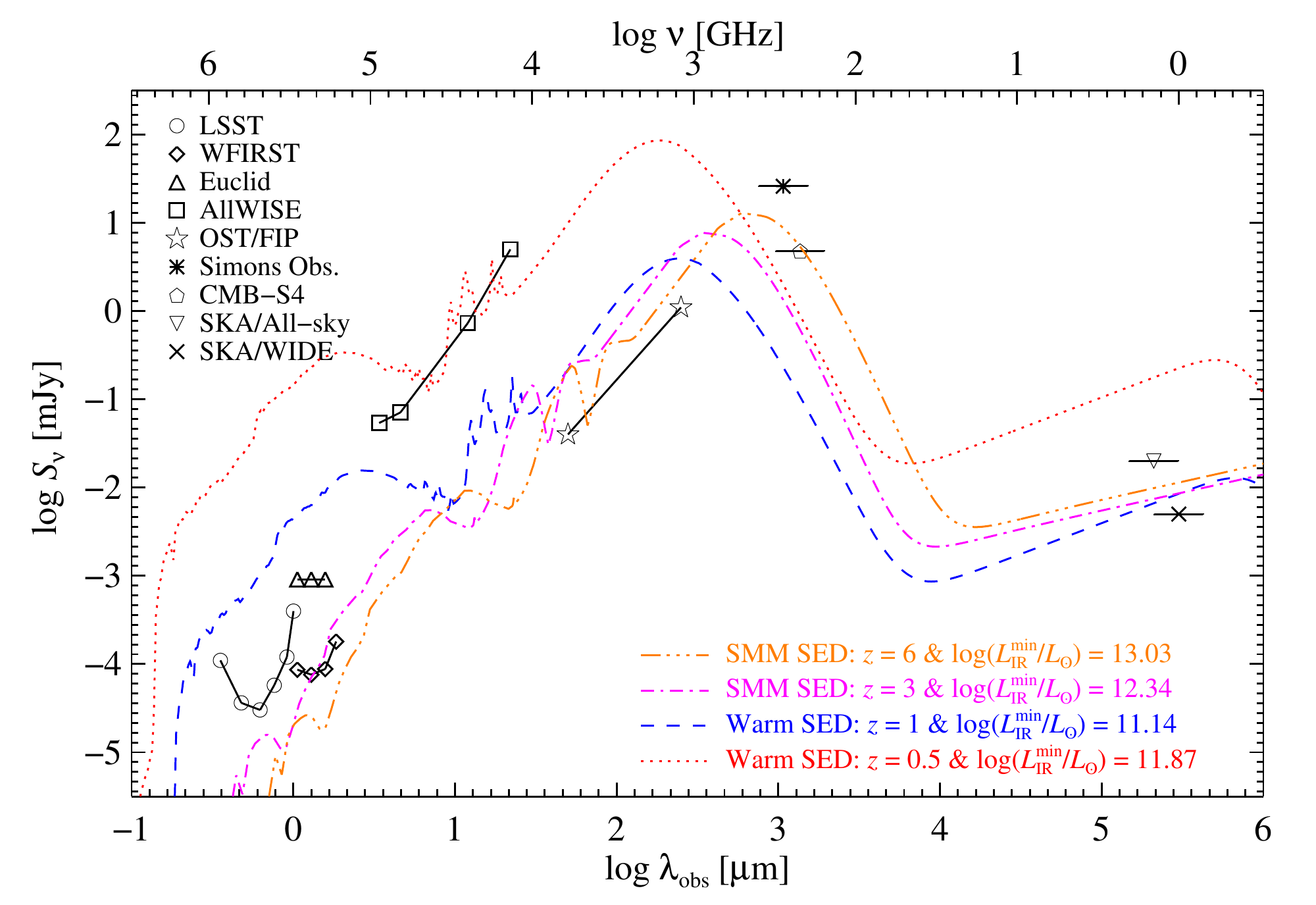}
\vspace{-4mm} \caption{Detection limits ($5\,\sigma$) of large area surveys at optical \citep{LSST2009},
near-IR \citep[\textit{Euclid} and \textit{WFIRST},][]{Laureijs2011, Spergel2015}, mid-IR
\citep[AllWISE,][and \url{http://wise2.ipac.caltech.edu/docs/release/allwise/expsup/sec2_3a.html}]{Cutri2013},
far-IR/submm \citep[OST/FIP, CMB-S4 and Simons Observatory,][]{Cooray2019, Abazajian2019, Ade2019}, and radio \citep[SKA,][]{PrandoniSeymour2015}
wavelengths compared with model SEDs of galaxies having the minimum $L_{\rm IR}$ detectable in lines at $z\,{=}\,0.5$,
1, 3, and 6. At the two lower redshifts we have adopted the ``warm'' (starburst) SED, while at the two
higher redshifts, we use the proto-spheroid SED \citep{Cai2013}.}

\label{fig:det_lim}
\end{figure}

\section{Comparison and synergies with other large-area surveys}\label{sect:comparison}

Large samples of strongly lensed and unlensed dusty galaxies are expected to
also be  obtained by next generation CMB experiments, like the ``CMB-S4''
\citep[ground-based;][CMB-S4;]{Abazajian2019} and the space-borne Probe of
Inflation and Cosmic Origins \citep[\textit{PICO};][]{Hanany2019}. However, the
depth of \textit{PICO} surveys is limited by confusion, due to the modest
telescope size (1.4\,m).

Confusion is a limiting factor also in the case of the 6-m telescopes used as
part  of CMB-S4, since they operate at mm wavelengths. Another practical limit
is set by atmospheric noise. If the effective depths of the CMB-S4 surveys are
similar to those of the SPT (a 10-m telescope), the detection of only a few
hundred strongly lensed galaxies at $z\,{\ge}\,6$ is expected. However the
CMB-S4 survey will provide deeper photometry at mm wavelengths, which is
important to quantify the cold dust emission.

A great advantage of a spectroscopic survey is the direct measurement of
redshifts,  while surveys of the continuum require a lengthy follow-up
programme that may be impractical for hundreds of thousands of optically very
faint or almost invisible
\citep[e.g.,][]{Dannerbauer2002,Younger2007,Dannerbauer2008,Wang2019} galaxies.
In addition, the target lines allow us to single out the star-formation
luminosity. It may be difficult to disentangle this from the AGN contribution
using broadband photometry alone \citep{Symeonidis2016,Symeonidis2017}.

As illustrated in Fig.~\ref{fig:det_lim}, there are important synergies with
large-area surveys at other wavelengths. For example, the Large Synoptic Survey
Telescope \citep[LSST;][]{LSST2009} will survey $20{,}000\,\hbox{deg}^2$ of the
sky in six photometric bands. The final coadded depths (point sources;
$5\sigma$) are $u= 26.3$, $g=27.5$, $r= 27.7$, $i=27.0$, $z=26.2$, and $y=24.9$
AB magnitudes.

\textit{Euclid} \citep{Laureijs2011}  will cover $15{,}000\,\hbox{deg}^2$ of the
sky to $Y$, $J$, and $H=24\,$mag. The \textit{Wide Field Infrared Survey Telescope}
\citep[\textit{WFIRST};][]{Spergel2015} will carry out large area, deep multi-filter
imaging surveys at high Galactic latitudes. As an example, we show in
Fig.~\ref{fig:det_lim} the expected depths for a nominal 2.5-year
survey of $2{,}500\,\hbox{deg}^2$ down to $J\sim 27$ AB mag. The AllWISE survey
\citep{Wright2010} has already provided shallower all-sky surveys at 3.4, 4.6, 12, and
$22\,\mu$m.

The Far-infrared Imager and Polarimeter (FIP) on the {\it Origins Space
Telescope\/}  \citep[OST;][]{Cooray2019} will deliver an ultra-wide-field
survey ($10{,}000\,\hbox{deg}^2$) at $250\,\mu$m down to the confusion limit
(1.1\,mJy) and a Wide survey ($500\,\hbox{deg}^2$) at $50\,\mu$m down to
$40\,\mu$Jy. The Simons Observatory \citep{Ade2019} Large Aperture Telescope
will survey 40\,\% of the sky with arc-minute resolution down to 26\,mJy at
280\,GHz. CMB-S4 \citep{Abazajian2019} will reach the confusion limit
($4.8\,$mJy at 220\,GHz) over a large fraction of the sky, and will cover
several frequency bands.

The Square Kilometer Array (SKA) ``all-sky'' survey will cover
$31{,}000\,\hbox{deg}^2$ at 1.4\,GHz down to $20\,\mu{\rm Jy}\,{\rm beam}^{-1}$ ($5\,\sigma$),
while the SKA/Wide survey at around 1\,GHz will reach $5\,\mu{\rm Jy}\,{\rm beam}^{-1}$
($5\,\sigma$) over $1{,}000\,\hbox{deg}^2$ \citep{PrandoniSeymour2015}.

As illustrated in Fig.~\ref{fig:det_lim}, the \textit{WFIRST}, the
\textit{Origins}/FIP and the SKA/Wide surveys are expected to detect
practically all the dusty galaxies seen in lines in our proposed spectroscopic
survey, although over limited sky areas. On the other hand,  the other surveys
mentioned above will also detect a substantial fraction of them, thus providing
complementary information on stellar and dust components.

\begin{figure}[htbp!]
\centering
\includegraphics[width=\linewidth]{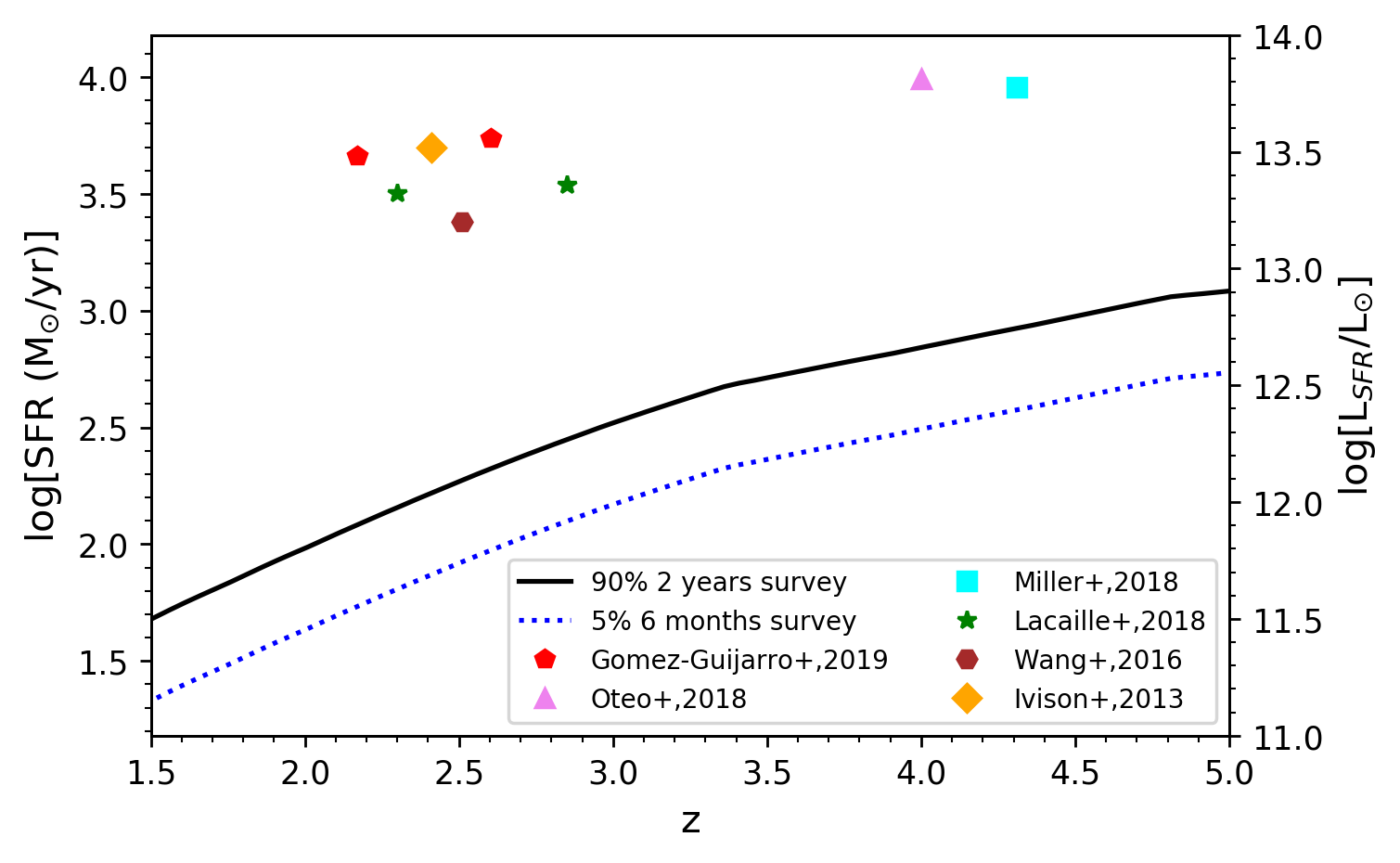}
\vspace{-4mm} \caption{SFRs of the spectroscopically confirmed, submm-bright proto-cluster cores discovered so far,
compared with the minimum SFRs detectable at $5\,\sigma$ by the ``all-sky'' (solid black line; 2 years,
90\,\% of the sky) and by the ``deep'' (dotted blue line; 5\,\% of the sky, six months) spectroscopic surveys.
The data points are: \citet{Ivison2013} at $z\,{=}\,2.41$; \citet{GomezGuijarro2019} at $z\,{=}\,2.171$ and $z\,{=}\,2.602$;
\citet{Wang2016} at $z\,{=}\,2.51$; \citet{Miller2018} at
$z\,{=}\,4.31$; \citet{Oteo2018} at $z\,{=}\,4.0$; and \citet{Lacaille2018} at $z \,{\simeq}\,2.85$ and ${\simeq}\,2.30$.}
\label{fig:proto_cores}
\end{figure}

\begin{figure}[htbp!]
\centering
\includegraphics[width=\linewidth]{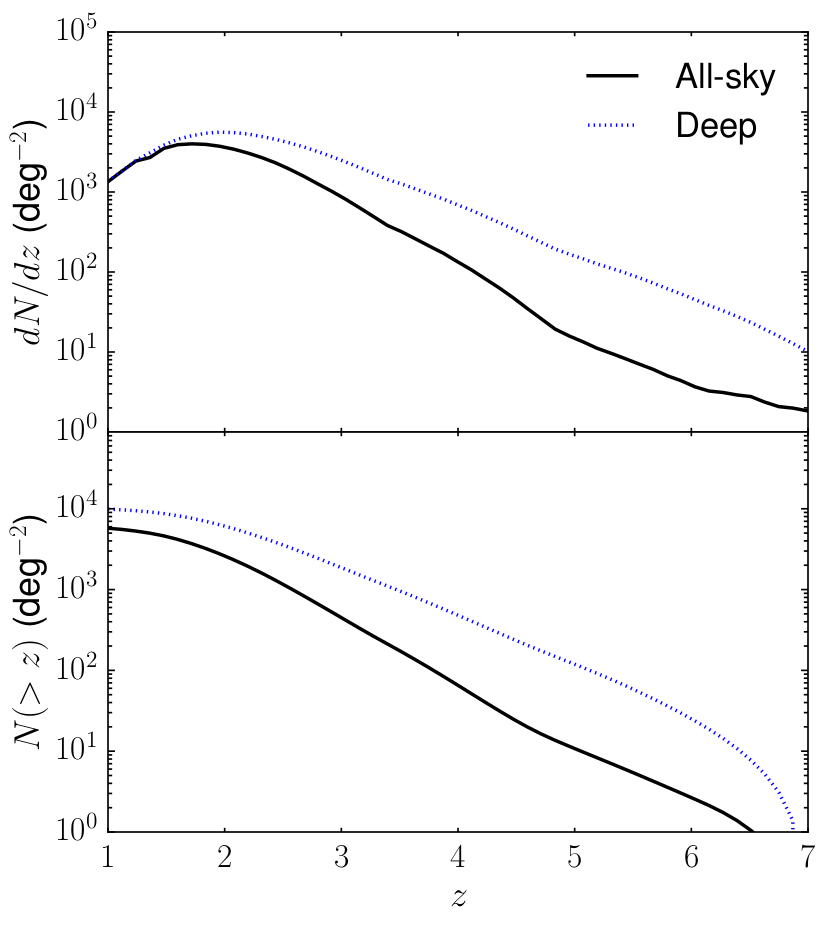}
\vspace{-4mm} \caption{Predicted differential and cumulative redshift distributions
(upper and lower panels, respectively) of proto-clusters detected in at least
one line by the ``all-sky'' survey (2\,yr, 90\,\% of the sky; solid black line) and by
the deep survey (6 months, 5\,\% of the sky; dotted blue line).
We expect the detection of tens of millions of proto-clusters at $z\simeq 2$ and of tens of thousands
of them at $z\simeq 6$.}
\label{fig:protocl_zdistr}
\end{figure}

\section{Revealing galaxy proto-clusters via dusty
starbursts}\label{sect:protocluster}

N-body simulations in the framework of the currently standard $\Lambda$CDM
cosmology have elucidated how primordial perturbations have grown into
collapsed halos distributed within a filamentary structure  \citep[the cosmic
web; e.g.][]{Springel2005, BoylanKolchin2009}.
However an observational validation of how these objects are assembled is still
missing. 
Understanding the full evolutionary history of present-day galaxy clusters is
of  fundamental importance for the observational validation of the formation
history of the most massive dark-matter halos, a crucial test of models for
structure formation, as well as for investigating the impact of environment on
the formation and evolution of galaxies \citep{KravtsovBorgani2012,
OverzierKashikawa2019, Dannerbauer2019b}.


To address the many still open questions on cluster formation and evolution we
need to follow all their evolutionary stages through cosmic time, starting from
their progenitors, galaxy clusters in formation, so-called ``proto-clusters''
\citep[for a review see][]{Overzier2016}. This needs a coordinated
multi-frequency effort. Cluster identification via classical methods
(optical/IR imaging and detection of X-ray or SZ signals from the hot ICM) has
been very effective at relatively low redshifts. Samples of SZ-selected
clusters have been recently extended to a few thousand objects, primarily
thanks to surveys with the \textit{Planck} satellite
\citep{PlanckCollaboration2016SZ}, the South Pole Telescope
\citep[SPT;][]{Bleem2019} and the Atacama Cosmology Telescope
\citep[ACT;][]{Hilton2018}. A few thousand clusters have also been detected in
X-rays \citep{Klein2019}. Both ICM-based techniques have yielded just a handful
of clusters above $z\,{\simeq}\,1.5$. The \textit{e-ROSITA} all-sky survey is
expected to boost the number of X-ray-detected clusters to around $10^5$, but
again few detections are expected at $z\,{\simgt}\,1.5$ \citep{Grandis2019}.
There is also a limit to detecting X-rays and the SZ effect at very high
redshifts because of the lack of virialised gas.

Galaxy cluster searches looking for overdensities of galaxies in large-area
optical/IR  surveys \citep[e.g.,][]{Oguri2018, WenHan2018, Gonzalez2019} are
generally limited to $z\,{\simlt}\,1.5$ due to the subtle density contrasts of
the object. Observations indicate that $z\,{\simeq}\,1.5$ corresponds to a
critical epoch in galaxy cluster evolution. At lower redshifts the global SFR
of galaxies is anti-correlated with local density \citep{Dressler1980,
Kauffmann2004, Lemaux2019}. However, the specific SFR (i.e., the SFR per unit
stellar mass) increases faster in clusters than in the field, catching up with
that in the field at $z\,{\simeq}\,1.5$ \citep{Alberts2014, Alberts2016,
Wagner2017}. Thus galaxy clusters selected through optical/IR observations are
rare at $z\,{\simgt}\,1.5$ \citep{Gobat2011}.

\begin{figure*}[htbp!]
\includegraphics[width=\textwidth]{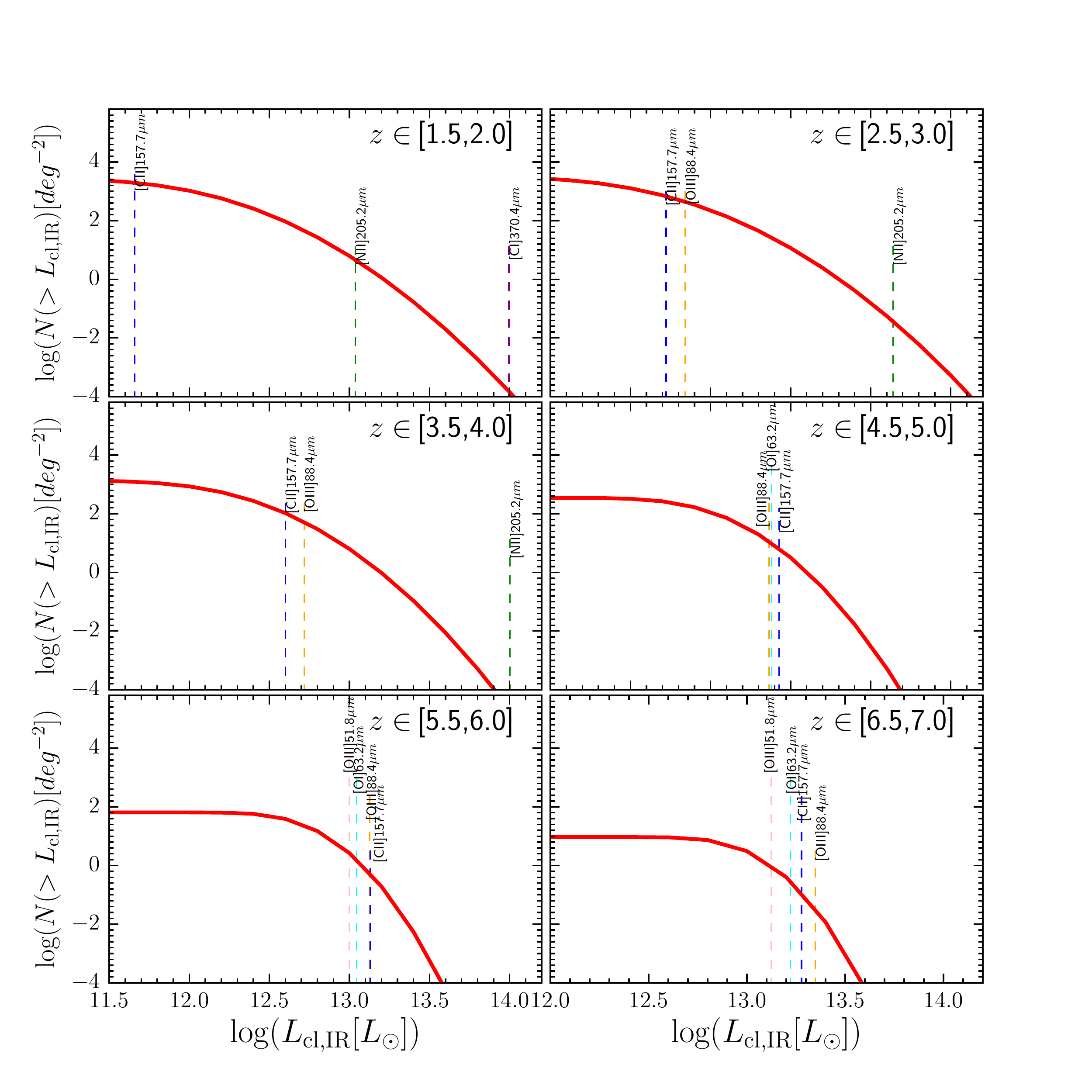}
\caption{Cumulative IR (8--$1000\,\mu$m) luminosity functions of proto-clusters
within $\delta z= 0.5$ at six redshifts. The predictions are based on the
model of \citet{Negrello2017protocl}.  The line luminosities corresponding to $L_{\rm IR}$ were
computed as described in the text. The vertical lines show the detection limits
for the brightest lines, assuming the instrument performances described
in the text. Such an instrument will detect strongly lensed galaxies (cf.\ Fig.\,\ref{fig:galLFs}) and proto-clusters of dusty galaxies all
the way out to the reionization redshift.
}
\label{fig:protoclusters}
\end{figure*}

Above $z\,{\simeq}\,1.5$, when a large fraction of member galaxies are in the
dust-obscured star-formation phase \citep[e.g.,][]{Dannerbauer2014,
Clements2016, Kato2016, Wagner2017, Nantais2017}, proto-cluster searches are
most conveniently carried out at submm wavelengths. \citet{Negrello2005}
predicted the detection by \textit{Planck} and \textit{Herschel} submm surveys
of unresolved intensity peaks made by the summed emission of dusty star-forming
high-$z$ proto-cluster members within the beam. Motivated by this work,
\citet{Planck2016highz} have reported the detection of 2151 proto-cluster
candidates over the cleanest 26\,\% of the sky -- these are unresolved sources
with ``red'' submm colours, consistent with redshifts $z\,{>}\,2$.

Follow-up observations of a subset of these sources with {\it Herschel}
\citep{PlanckHerschel2015} have shown that, apart from a tiny fraction of
strongly lensed galaxies \citep[the GEMS sample, see][]{Canameras2015}, they
are associated with overdensities of dusty star-forming galaxies. However,
\citet{Negrello2017protocl} argued that most of them are probably not
individual proto-clusters, but could be made of physically unrelated high-$z$
structures falling by chance within the relatively large \textit{Planck} beam
($\hbox{FWHM}\,{\simeq}\,5^\prime$). This has been confirmed by spectroscopic
measurements and/or photometric redshift estimates for two objects, showing
that each consists of two independent overdensities along the line of sight
\citep{FloresCacho2016, Kneissl2019}. 


The confusion problem (blending of independent proto-clusters along the line of
sight) affecting \textit{Planck} surveys is due to its poor angular resolution.
The instrument considered here has angular resolution $\le 1'$ at $\nu \ge
380\,$GHz (cf. Table~\ref{tab:sensitivities}). This is the ideal angular
resolution for searches of submm-bright proto-clusters. The angular correlation
function of faint submm galaxies measured by \citet{Chen2016} in photometric
redshift bins up to $z\,{\simeq}\,5$ showed that the 1-halo component,
corresponding to the source distribution within the proto-cluster halo,
dominates at $\theta\,{\simlt}\,1^\prime$. This is in keeping with the results
of \citet{Alberts2014} on 274 clusters with $0.3\,{\le}\,z\,{\le}\,1.5$. They
found that the density of IR-emitting cluster members clearly exceeds that of
the background field level only within $0.5\,$Mpc of the cluster centre. Such a
linear scale corresponds to an angular scale of about $1'$ at redshifts in the
range 1.5--2.5.

The clustering measurements by \citet{Chen2016} also gave clear indications of
a steepening of the correlation function on arcsec angular scales.
\citet{Negrello2017protocl} modelled this in terms of an additional component
with a spatial correlation function $\xi(r)\propto r^{-3.3}$. This component,
which dominates for $r\,{\simlt}\,50\,$kpc, may represent the cluster core. 
Submm-bright proto-cluster cores, with linear diameters of order $100\,$kpc,
have been discovered by \citet{Ivison2013, Ivison2019} at $z=2.41$,
\citet{GomezGuijarro2019} at $z\,{=}\,2.2$ and $z\,{}\,2.6$, \citet{Wang2016}
at $z\,{=}\,2.5$, and \citet{Miller2018} at $z\,{=}\,4.3$. \citet{Oteo2018}
reported the identification of a dusty proto-cluster core at $z\,{=}\,4.0$,
whose member galaxies are distributed over a $260\,\hbox{kpc}\times
310\,\hbox{kpc}$ region. Additionally \citet{Lacaille2018} spectroscopically
confirmed the detection of two submm proto-clusters at $z\,{\simeq}\,2.8$ and
${\simeq}\,2.3$, with $1'$ angular radius.

The spectroscopically confirmed proto-cluster \textit{cores} listed above are
unresolved at the resolution of the instrument, except perhaps at the highest
frequencies, where the maps can be degraded to arcmin resolution. The summed
IR-luminosities of detected members of the spectroscopically confirmed high-$z$
proto-clusters (lower limits to the proto-cluster total IR luminosities) range
from a few to $\hbox{several}\times 10^{13}\,{\rm L}_\odot$ and are all well
above the line-detection limits (see Fig.~\ref{fig:proto_cores}) of our
proposed surveys.

High-$z$ proto-clusters with enhanced star formation on larger (few to several
Mpc) scales have also been reported by \citet{Dannerbauer2014},
\citet{Casey2015}, \citet{Kato2016}, \citet{Hung2016}, and \citet{Umehata2019}.
\textit{All} proto-clusters with similar properties will be detected by our
spectroscopic survey.

As illustrated in Figs.~\ref{fig:protocl_zdistr} and \ref{fig:protoclusters},
the instrument considered here will detect millions of proto-clusters at the
peak of cosmic star-formation activity ($z=2$--3).  For a large fraction of
proto-clusters, especially at $z\ge 2.4$, at least two lines will be detected,
allowing a solid redshift determination without requiring follow-up
observations.

\section{Comparison with other large-area cluster
surveys}\label{sect:cluster_surveys}

No other foreseen survey can do anything similar to the future spectroscopic
satellite. {\it Euclid\/}'s ``Wide'' survey over $15{,}000\,\hbox{deg}^2$ is
expected to detect galaxy clusters and proto-clusters only out to
$z\,{\simeq}\,2$. The estimated surface density of clusters detected at
$5\,\sigma$ in the redshift range $1.5\,{\le}\,z\,{\le}\,2$ is $\simeq
0.19\,\hbox{deg}^{-2}$ \citep[][see their figure~3]{Sartoris2016}.

A survey of $8{,}300\,\hbox{deg}^2$ by \textit{WFIRST}  \citep{Spergel2015} is
expected to detect 20{,}000 clusters with masses $M_{\rm cl}\ge 7.4\times
10^{13}\,{\rm M}_\odot$ at $z\,{=}\,1.5$--2 and 2{,}800 clusters at
$z\,{=}\,2$--2.5 for the same masses. The corresponding surface densities are
of 2.4 and $0.34\,\hbox{deg}^{-2}$, respectively.

For comparison, we expect around $2000\,\hbox{deg}^{-2}$ ($1300\,
\hbox{deg}^{-2}$) detections in the [{\sc Cii}]\,157.7\,$\mu$m line only and
$\simeq 5\,\hbox{deg}^{-2}$ ($\simeq 0.4\,\hbox{deg}^{-2}$) detections in two
lines ([{\sc Cii}]\,157.7\,$\mu$m and [{\sc Nii}]\,205.18\,$\mu$m) at $1.5\,{\le}\,z\,{\le}\,2$
($2.0\,{\le}\,z\,{\le}\,2.5$), as shown in Fig.~\ref{fig:protoclusters}.

Cluster searches with ground-based, large-area optical  surveys like LSST
\citep{LSST2009} and the Javalambre Physics of the Accelerating Universe
Astrophysical Survey \citep[J-PAS;][]{Benitez2014} are limited to
$z\,{<}\,1.5$. As mentioned above, the X-ray cluster survey of
\textit{e-ROSITA} is also limited to $z\,{\simlt}\,1.5$ \citep{Grandis2019}.

The most extensive optical search for high-$z$ proto-clusters is being
conducted using a wide-field survey with the Hyper Suprime-Cam (HSC)
instrument mounted at the prime focus of the Subaru telescope. The wide HSC
survey will cover $1{,}400\,\hbox{deg}^2$. \citet{Toshikawa2018} carried
out a blind search for proto-clusters at $z\,{\simeq}\,3.8$ over an area
of $121\,\hbox{deg}^2$ using colour criteria to select galaxies at
$z\,{\simeq}\,3.3$--4.2. They found 216 overdensities that were significant
at the ${>}\,4\,\sigma$ level. The estimated comoving density of these objects
is $1.4\times 10^{-7}\,\hbox{Mpc}^{-3}$ for the cosmology adopted here.
At higher redshifts \citet{Higuchi2019} reported 14 and 26 proto-cluster
candidates at $z\,{=}\,5.7$ and $z\,{=}\,6.6$, covering an area of 14 and
$16\,\hbox{deg}^2$, respectively.

\citet{Kubo2019} reported statistical evidence of submm emission from HSC
candidate proto-clusters in \textit{Planck} high-frequency maps, confirming the
ubiquity of intense star formation in high-$z$ proto-clusters. The estimated
average IR luminosity and SFR are well above the detection limits of
the proposed spectroscopic survey. 

{At $z\,{=}\,3.75$, i.e., close to the redshift of the optical search by
\citet{Toshikawa2018}, the spectroscopic survey will detect
$100\,\hbox{deg}^{-2}$ proto-clusters in the [{\sc Cii}]\,157.7\,$\mu$m line
and $50\,\hbox{deg}^{-2}$ also in the [{\sc Oiii}]\,88.36\,$\mu$m line. The
corresponding comoving volume densities are 1--$2\times
10^{-5}\,\hbox{Mpc}^{-3}$, i.e., about two orders of magnitude higher than that
achieved by \citet{Toshikawa2018}. As pointed out by the latter authors,
overdensity searches via optical imaging surveys are expected to be highly
incomplete (completeness around 6\,\%) because most overdensities are swamped
by projection effects. On the contrary, the statistical detection on
\textit{Planck} maps of large SFRs in optically-selected proto-clusters
suggests a high completeness level for the planned spectroscopic survey.}



It is important to stress that the proposed instrument can measure both the SFR
and the SZ effect in clusters. It will thus simultaneously probe the evolution
of star formation in dense environment and of the hot IGM, allowing us to
investigate the relationship between the two aspects of cluster evolution. It
will thereby provide key data on the still unexplored, crucial transition
period at $z\,{\simeq}\,1.5$--2 when cluster galaxies were vigorously forming
stars and at the same time the hot IGM was taking root.

Figure~\ref{fig:protoclusters} shows that the spectroscopic survey will detect
enough proto-clusters within relatively narrow redshift bins to accurately
measure their two-point spatial correlation, $\xi(r)$, as a function of
redshift out to $z\,{\simeq}\,6$ and, at fixed $z$, as a function of $L_{\rm
IR}$ (or equivalently of SFR). From the function $\xi(r)$ we can infer the
effective halo mass \citep{Sheth2001}, so that the observed proto-cluster
abundance provides constraints on the evolution of the high-mass tail of the
halo mass function and on the relationship between halo mass and $L_{\rm IR}$
or SFR.

Note that the abundance and clustering of galaxy clusters are both probes of
large-scale structure growth. Their determination as a function of mass and
redshift enables us to constrain cosmological parameters primarily through the
linear growth rate of perturbations. This has been proven to be competitive
with and complementary to other probes
\citep[e.g.,][]{PlanckCollaboration2016SZcosm, deHaan2016, Sridhar2017}.

At lower redshifts, the comparison of cluster samples detected via star
formation and via the SZ effect will quantify the history of star-formation
quenching in galaxy clusters, a still open issue \citep{Boselli2016,
RodriguezMunoz2019}. In particular, we will learn about the quenching timescale
at different redshifts and for different cluster halo masses, inferred from the
amplitude of the SZ effect, and of the ICM density inferred from X-ray
observations, e.g., from \textit{e-ROSITA}. This is key information for
identifying the mechanism(s) responsible for the environmental quenching. 

\section{Conclusions}\label{sect:conclusions}

The high-sensitivity spectroscopic and imaging surveys carried out by the space
mission proposed by \citet{Delabrouille2019} will revolutionize our view of
galaxy evolution and of the growth of galaxy clusters.

Spectroscopy enables full exploitation of the sensitivity of present-day
instrumentation, bypassing the confusion limits that severely constrain the
depth of submm surveys with telescopes of the 3-m class, like that of the
\textit{Herschel} Observatory.

The high luminosity of far-IR lines such as [{\sc Cii}]\,157.7\,$\mu$m, [{\sc
Oiii}]\,88.36\,$\mu$m, [{\sc Oi}]63.18$\mu$m and [{\sc Oiii}]\,51.81\,$\mu$m,
makes it possible to detect, at $z\,{>}\,1$, star-forming galaxies with SFRs
about one order of magnitude lower that those reached by the
\textit{Herschel}-ATLAS survey. Moreover, we have argued that these
fine-structure lines track the full SFR, not only the dust-obscured fraction
measured by the far-IR/submm continuum.

Most importantly, the spectroscopic survey directly provides the 3D
distribution of star-forming galaxies all the way through the reionization
epoch. Many tens of millions of galaxies will be detected over the redshift
range $1\,{<}\,z\,{<}\,3$ (Fig.~\ref{fig:gal_zdistr}) where the cosmic SFR
peaks. SFRs well below those of typical galaxies at these redshifts will be
reached (Figs.~\ref{fig:LIR_lim} and \ref{fig:galLFs}). For example, at
$z\,{=}\,2$ the [{\sc Cii}]\,157.7\,$\mu$m line will allow the detection of
SFRs down to about $60\,{\rm M}_\odot\,\hbox{yr}^{-1}$ by the ``all-sky''
survey and down to $30\,{\rm M}_\odot\,\hbox{yr}^{-1}$ for the ``deep'' survey.
Our conservative estimates, that do not allow for the possibility of a
top-heavier IMF at high $z$ indicated by theoretical arguments and by some
observational results \citep[see][for a discussion]{Cai2019}, yield the
detection of thousands of galaxies at $z\,{\simeq}\, 6$ and of several tens at
$z\simeq 8$.

This outcome is far better than can be achieved by existing and forthcoming
continuum surveys, which, moreover, require time-consuming redshift follow-up
programmes that are impractical for millions of optically very faint sources.

The spectroscopic survey will detect millions of strongly lensed galaxies,
which  dominate the bright tails of the high-$z$ SFR functions, above
uncorrected SFRs of a few thousand ${\rm M}_\odot\,\hbox{yr}^{-1}$.
Additionally, strong lensing will allow us to probe SFR functions below the
nominal detection limit.

The brightest strongly lensed galaxies are obvious targets for follow-up
observations  with high-resolution instruments like ALMA or {\it JWST}. The
combination of their extreme luminosity and of the stretching of their images
offers a unique possibility of peering into the internal structure of high-$z$
galaxies down to scales of tens of parsecs, comparable to or smaller than the
size of giant molecular clouds in the Milky Way. Data at this resolution are
the only way to directly investigate the complex physical processes driving the
early evolution of galaxies.

The arcmin (or better) angular resolution at submm wavelengths of the proposed
survey is ideal  to detect proto-cluster cores out to $z\,{\simeq}\,7$. The
summed line emission of star-forming member galaxies within the beam of the
instrument makes these objects produce the brightest intensity peaks at each
redshift. Proto-clusters can thus be detected independently of whether they
contain the hot IGM that would make than detectable in X-rays or via the SZ
effect.

We predict the spectroscopic detection of millions of proto-clusters at
$z\,{>}\,1.5$,  i.e., in the redshift range hardly accessible to classical
cluster-detection methods. This will provide a complete view of the
star-formation history in dense environments. The statistical detection of
intense star formation in optically-selected high-$z$ proto-clusters
\citep{Kubo2019} suggests that the planned spectroscopic survey will have a
high completeness level.

Enough proto-clusters will be detected within relatively narrow redshift bins
to accurately measure their clustering $\xi(r,z)$ out to $z\,{\simeq}\,6$ and,
at fixed $z$, as a function of their SFR. Since $\xi(r)$ allows us to estimate
the halo mass, we can derive the first observational constraints on the
evolution of the high-mass tail of the halo mass function and on the
relationship between halo mass and SFR.

At lower redshifts ($z\,{\simlt}\,1.5$--2), the mission will simultaneously
measure  the SFR and the SZ effect for early structures. In combination with
X-ray data from \textit{e-Rosita} and with data from optical surveys measuring
stellar masses, this will elucidate several key issues, e.g., the transition
from the active star-forming to passive evolution phases of cluster members,
the origin of the hot IGM, and the mechanisms responsible for the environmental
quenching of star formation.

In all these areas, the proposed spectroscopic survey transcends any other
foreseen project.

\begin{acknowledgements}
We are grateful to the referee for useful comments. H.D. acknowledges financial
support from the Spanish Ministry of Science, Innovation,  and Universities
(MICIU) under the 2014 Ram\'on y Cajal program RYC-2014-15686 and
AYA2017-84061-P, the latter being co-financed by FEDER (European Regional
Development Funds). M.B. acknowledges partial financial support from the
Italian Ministero dell'Istruzione, Universit\`{a} e Ricerca through the grant
`Progetti Premiali 2012 -- iALMA' (CUP C52I13000140001) and from INAF under
PRIN SKA/CTA FORECaST. Z.Y.C. is supported by the National Science Foundation
of China (grant No. 11890693). M.N. acknowledges financial support from the
European Union's Horizon 2020 research and innovation programme under the Marie
Sk{\l}odowska-Curie grant agreement No. 707601.
\end{acknowledgements}

\bibliographystyle{pasa-mnras}
\bibliography{lines}

\begin{thebibliography}{}
\makeatletter
\relax
\def\mn@urlcharsother{\let\do\@makeother \do\$\do\&\do\#\do\^\do\_\do\%\do\~}
\definecolor{darkblue}{rgb}{0,0,0.597656}
\def\mndoi{\begingroup\mn@urlcharsother \@ifnextchar [ {\mndoi@} {\mndoi@[]}}
\def\mndoi@[#1]#2{\def\@tempa{#1}\ifx\@tempa\@empty \href
  {http://dx.doi.org/#2} {\textcolor{darkblue}{doi:#2}}\else \href
  {http://dx.doi.org/#2} {\textcolor{darkblue}{#1}}\fi \endgroup}
\def\mn@eprint#1#2{\mn@eprint@#1:#2::\@nil}
\def\mn@eprint@arXiv#1{\href {http://arxiv.org/abs/#1} {{\tt arXiv:#1}}}
\def\mn@eprint@dblp#1{\href {http://dblp.uni-trier.de/rec/bibtex/#1.xml}
  {dblp:#1}}
\def\mn@eprint@#1:#2:#3:#4\@nil{\def\@tempa {#1}\def\@tempb {#2}\def\@tempc
  {#3}\ifx \@tempc \@empty \let \@tempc \@tempb \let \@tempb \@tempa \fi \ifx
  \@tempb \@empty \def\@tempb {arXiv}\fi \@ifundefined
  {mn@eprint@\@tempb}{\@tempb:\@tempc}{\expandafter \expandafter \csname
  mn@eprint@\@tempb\endcsname \expandafter{\@tempc}}}

\bibitem[\protect\citeauthoryear{{Abazajian} et~al.,}{{Abazajian}
  et~al.}{2019}]{Abazajian2019}
{Abazajian} K.,  et~al., 2019, arXiv e-prints, \href
  {https://ui.adsabs.harvard.edu/abs/2019arXiv190704473A} {p. arXiv:1907.04473}

\bibitem[\protect\citeauthoryear{{Ade} et~al.,}{{Ade} et~al.}{2019}]{Ade2019}
{Ade} P.,  et~al., 2019, \mndoi [\jcap] {10.1088/1475-7516/2019/02/056}, \href
  {https://ui.adsabs.harvard.edu/abs/2019JCAP...02..056A} {2019, 056}

\bibitem[\protect\citeauthoryear{{Alberts} et~al.,}{{Alberts}
  et~al.}{2014}]{Alberts2014}
{Alberts} S.,  et~al., 2014, \mndoi [\mnras] {10.1093/mnras/stt1897}, \href
  {http://adsabs.harvard.edu/abs/2014MNRAS.437..437A} {437, 437}

\bibitem[\protect\citeauthoryear{{Alberts} et~al.,}{{Alberts}
  et~al.}{2016}]{Alberts2016}
{Alberts} S.,  et~al., 2016, \mndoi [\apj] {10.3847/0004-637X/825/1/72}, \href
  {http://adsabs.harvard.edu/abs/2016ApJ...825...72A} {825, 72}

\bibitem[\protect\citeauthoryear{{Bakx} et~al.,}{{Bakx}
  et~al.}{2018}]{Bakx2018}
{Bakx} T. J.~L.~C.,  et~al., 2018, \mndoi [\mnras] {10.1093/mnras/stx2267},
  \href {https://ui.adsabs.harvard.edu/abs/2018MNRAS.473.1751B} {473, 1751}

\bibitem[\protect\citeauthoryear{{Barger}, {Cowie}, {Sanders}, {Fulton},
  {Taniguchi}, {Sato}, {Kawara}  \& {Okuda}}{{Barger}
  et~al.}{1998}]{Barger1998}
{Barger} A.~J.,  {Cowie} L.~L.,  {Sanders} D.~B.,  {Fulton} E.,  {Taniguchi}
  Y.,  {Sato} Y.,  {Kawara} K.,   {Okuda} H.,  1998, \mndoi [\nat]
  {10.1038/28338}, \href
  {https://ui.adsabs.harvard.edu/abs/1998Natur.394..248B} {394, 248}

\bibitem[\protect\citeauthoryear{{Barger}, {Cowie}, {Smail}, {Ivison}, {Blain}
  \& {Kneib}}{{Barger} et~al.}{1999}]{Barger1999}
{Barger} A.~J.,  {Cowie} L.~L.,  {Smail} I.,  {Ivison} R.~J.,  {Blain} A.~W.,
  {Kneib} J.~P.,  1999, \mndoi [\aj] {10.1086/300890}, \href
  {https://ui.adsabs.harvard.edu/abs/1999AJ....117.2656B} {117, 2656}

\bibitem[\protect\citeauthoryear{{Basu} et~al.,}{{Basu}
  et~al.}{2019}]{Basu2019}
{Basu} K.,  et~al., 2019, arXiv e-prints, \href
  {https://ui.adsabs.harvard.edu/abs/2019arXiv190901592B} {p. arXiv:1909.01592}

\bibitem[\protect\citeauthoryear{{Baugh}, {Lacey}, {Frenk}, {Granato}, {Silva},
  {Bressan}, {Benson}  \& {Cole}}{{Baugh} et~al.}{2005}]{Baugh2005}
{Baugh} C.~M.,  {Lacey} C.~G.,  {Frenk} C.~S.,  {Granato} G.~L.,  {Silva} L.,
  {Bressan} A.,  {Benson} A.~J.,   {Cole} S.,  2005, \mndoi [\mnras]
  {10.1111/j.1365-2966.2004.08553.x}, \href
  {https://ui.adsabs.harvard.edu/abs/2005MNRAS.356.1191B} {356, 1191}

\bibitem[\protect\citeauthoryear{{Benitez} et~al.,}{{Benitez}
  et~al.}{2014}]{Benitez2014}
{Benitez} N.,  et~al., 2014, arXiv e-prints, \href
  {https://ui.adsabs.harvard.edu/abs/2014arXiv1403.5237B} {p. arXiv:1403.5237}

\bibitem[\protect\citeauthoryear{{Bernardi}, {Shankar}, {Hyde}, {Mei},
  {Marulli}  \& {Sheth}}{{Bernardi} et~al.}{2010}]{Bernardi2010}
{Bernardi} M.,  {Shankar} F.,  {Hyde} J.~B.,  {Mei} S.,  {Marulli} F.,
  {Sheth} R.~K.,  2010, \mndoi [\mnras] {10.1111/j.1365-2966.2010.16425.x},
  \href {https://ui.adsabs.harvard.edu/abs/2010MNRAS.404.2087B} {404, 2087}

\bibitem[\protect\citeauthoryear{{Bethermin} et~al.,}{{Bethermin}
  et~al.}{2020}]{Bethermin2020}
{Bethermin} M.,  et~al., 2020, arXiv e-prints, \href
  {https://ui.adsabs.harvard.edu/abs/2020arXiv200200962B} {p. arXiv:2002.00962}

\bibitem[\protect\citeauthoryear{{Biggs}, {Younger}  \& {Ivison}}{{Biggs}
  et~al.}{2010}]{Biggs2010}
{Biggs} A.~D.,  {Younger} J.~D.,   {Ivison} R.~J.,  2010, \mndoi [\mnras]
  {10.1111/j.1365-2966.2010.17120.x}, \href
  {https://ui.adsabs.harvard.edu/abs/2010MNRAS.408..342B} {408, 342}

\bibitem[\protect\citeauthoryear{{Blain} \& {Longair}}{{Blain} \&
  {Longair}}{1993}]{BlainLongair1993}
{Blain} A.~W.,  {Longair} M.~S.,  1993, \mndoi [\mnras]
  {10.1093/mnras/264.2.509}, \href
  {https://ui.adsabs.harvard.edu/abs/1993MNRAS.264..509B} {264, 509}

\bibitem[\protect\citeauthoryear{{Bleem} et~al.,}{{Bleem}
  et~al.}{2020}]{Bleem2019}
{Bleem} L.~E.,  et~al., 2020, \mndoi [\apjs] {10.3847/1538-4365/ab6993}, \href
  {https://ui.adsabs.harvard.edu/abs/2020ApJS..247...25B} {247, 25}

\bibitem[\protect\citeauthoryear{{Bonato} et~al.,}{{Bonato}
  et~al.}{2019}]{Bonato2019}
{Bonato} M.,  et~al., 2019, \mndoi [\pasa] {10.1017/pasa.2019.8}, \href
  {https://ui.adsabs.harvard.edu/abs/2019PASA...36...17B} {36, e017}

\bibitem[\protect\citeauthoryear{{Boselli} et~al.,}{{Boselli}
  et~al.}{2016}]{Boselli2016}
{Boselli} A.,  et~al., 2016, \mndoi [\aap] {10.1051/0004-6361/201629221}, \href
  {https://ui.adsabs.harvard.edu/abs/2016A&A...596A..11B} {596, A11}

\bibitem[\protect\citeauthoryear{{Bothwell} et~al.,}{{Bothwell}
  et~al.}{2013}]{Bothwell2013}
{Bothwell} M.~S.,  et~al., 2013, \mndoi [\mnras] {10.1093/mnras/sts562}, \href
  {https://ui.adsabs.harvard.edu/abs/2013MNRAS.429.3047B} {429, 3047}

\bibitem[\protect\citeauthoryear{{Boylan-Kolchin}, {Springel}, {White},
  {Jenkins}  \& {Lemson}}{{Boylan-Kolchin} et~al.}{2009}]{BoylanKolchin2009}
{Boylan-Kolchin} M.,  {Springel} V.,  {White} S. D.~M.,  {Jenkins} A.,
  {Lemson} G.,  2009, \mndoi [\mnras] {10.1111/j.1365-2966.2009.15191.x}, \href
  {https://ui.adsabs.harvard.edu/abs/2009MNRAS.398.1150B} {398, 1150}

\bibitem[\protect\citeauthoryear{{Ca{\~n}ameras} et~al.,}{{Ca{\~n}ameras}
  et~al.}{2015}]{Canameras2015}
{Ca{\~n}ameras} R.,  et~al., 2015, \mndoi [\aap] {10.1051/0004-6361/201425128},
  \href {https://ui.adsabs.harvard.edu/abs/2015A&A...581A.105C} {581, A105}

\bibitem[\protect\citeauthoryear{{Ca{\~n}ameras}, {Nesvadba}  \&
  {Kneissl}}{{Ca{\~n}ameras} et~al.}{2017}]{Canameras2017ALMA}
{Ca{\~n}ameras} R.,  {Nesvadba} N.,   {Kneissl} R. e.~a.,  2017, \mndoi [\aap]
  {10.1051/0004-6361/201630186}, \href
  {http://adsabs.harvard.edu/abs/2017A\%A...604A.117C} {604, A117}

\bibitem[\protect\citeauthoryear{{Ca{\~n}ameras} et~al.,}{{Ca{\~n}ameras}
  et~al.}{2018}]{Canameras2018outflow}
{Ca{\~n}ameras} R.,  et~al., 2018, \mndoi [\aap] {10.1051/0004-6361/201833679},
  \href {http://adsabs.harvard.edu/abs/2018A\%A...620A..60C} {620, A60}

\bibitem[\protect\citeauthoryear{{Cai} et~al.,}{{Cai} et~al.}{2013}]{Cai2013}
{Cai} Z.-Y.,  et~al., 2013, \mndoi [\apj] {10.1088/0004-637X/768/1/21}, \href
  {http://adsabs.harvard.edu/abs/2013ApJ...768...21C} {768, 21}

\bibitem[\protect\citeauthoryear{{Cai}, {Lapi}, {Bressan}, {De Zotti},
  {Negrello}  \& {Danese}}{{Cai} et~al.}{2014}]{Cai2014}
{Cai} Z.-Y.,  {Lapi} A.,  {Bressan} A.,  {De Zotti} G.,  {Negrello} M.,
  {Danese} L.,  2014, \mndoi [\apj] {10.1088/0004-637X/785/1/65}, \href
  {https://ui.adsabs.harvard.edu/abs/2014ApJ...785...65C} {785, 65}

\bibitem[\protect\citeauthoryear{{Cai}, {De Zotti}  \& {Bonato}}{{Cai}
  et~al.}{2020}]{Cai2019}
{Cai} Z.-Y.,  {De Zotti} G.,   {Bonato} M.,  2020, \mndoi [\apj]
  {10.3847/1538-4357/ab7231}, \href
  {https://ui.adsabs.harvard.edu/abs/2019arXiv191006970C} {891, 74}

\bibitem[\protect\citeauthoryear{{Carniani} et~al.,}{{Carniani}
  et~al.}{2017}]{Carniani2017}
{Carniani} S.,  et~al., 2017, \mndoi [\aap] {10.1051/0004-6361/201630366},
  \href {https://ui.adsabs.harvard.edu/abs/2017A&A...605A..42C} {605, A42}

\bibitem[\protect\citeauthoryear{{Carniani} et~al.,}{{Carniani}
  et~al.}{2018}]{Carniani2018}
{Carniani} S.,  et~al., 2018, \mndoi [\mnras] {10.1093/mnras/sty1088}, \href
  {https://ui.adsabs.harvard.edu/abs/2018MNRAS.478.1170C} {478, 1170}

\bibitem[\protect\citeauthoryear{{Casey} et~al.,}{{Casey}
  et~al.}{2012a}]{Casey2012a}
{Casey} C.~M.,  et~al., 2012a, \mndoi [\apj] {10.1088/0004-637X/761/2/139},
  \href {https://ui.adsabs.harvard.edu/abs/2012ApJ...761..139C} {761, 139}

\bibitem[\protect\citeauthoryear{{Casey} et~al.,}{{Casey}
  et~al.}{2012b}]{Casey2012b}
{Casey} C.~M.,  et~al., 2012b, \mndoi [\apj] {10.1088/0004-637X/761/2/140},
  \href {https://ui.adsabs.harvard.edu/abs/2012ApJ...761..140C} {761, 140}

\bibitem[\protect\citeauthoryear{{Casey}, {Narayanan}  \& {Cooray}}{{Casey}
  et~al.}{2014}]{Casey2014}
{Casey} C.~M.,  {Narayanan} D.,   {Cooray} A.,  2014, \mndoi [\physrep]
  {10.1016/j.physrep.2014.02.009}, \href
  {https://ui.adsabs.harvard.edu/abs/2014PhR...541...45C} {541, 45}

\bibitem[\protect\citeauthoryear{{Casey} et~al.,}{{Casey}
  et~al.}{2015}]{Casey2015}
{Casey} C.~M.,  et~al., 2015, \mndoi [\apjl] {10.1088/2041-8205/808/2/L33},
  \href {https://ui.adsabs.harvard.edu/abs/2015ApJ...808L..33C} {808, L33}

\bibitem[\protect\citeauthoryear{{Chapman}, {Blain}, {Smail}  \&
  {Ivison}}{{Chapman} et~al.}{2005}]{Chapman2005}
{Chapman} S.~C.,  {Blain} A.~W.,  {Smail} I.,   {Ivison} R.~J.,  2005, \mndoi
  [\apj] {10.1086/428082}, \href
  {https://ui.adsabs.harvard.edu/abs/2005ApJ...622..772C} {622, 772}

\bibitem[\protect\citeauthoryear{{Chen} et~al.,}{{Chen}
  et~al.}{2016}]{Chen2016}
{Chen} C.-C.,  et~al., 2016, \mndoi [\apj] {10.3847/0004-637X/831/1/91}, \href
  {https://ui.adsabs.harvard.edu/abs/2016ApJ...831...91C} {831, 91}

\bibitem[\protect\citeauthoryear{{Chluba} et~al.,}{{Chluba}
  et~al.}{2019}]{Chluba2019}
{Chluba} J.,  et~al., 2019, arXiv e-prints, \href
  {https://ui.adsabs.harvard.edu/abs/2019arXiv190901593C} {p. arXiv:1909.01593}

\bibitem[\protect\citeauthoryear{{Clements} et~al.,}{{Clements}
  et~al.}{2016}]{Clements2016}
{Clements} D.~L.,  et~al., 2016, \mndoi [\mnras] {10.1093/mnras/stw1224}, \href
  {https://ui.adsabs.harvard.edu/abs/2016MNRAS.461.1719C} {461, 1719}

\bibitem[\protect\citeauthoryear{{Cooke} et~al.,}{{Cooke}
  et~al.}{2018}]{Cooke2018}
{Cooke} E.~A.,  et~al., 2018, \mndoi [\apj] {10.3847/1538-4357/aac6ba}, \href
  {https://ui.adsabs.harvard.edu/abs/2018ApJ...861..100C} {861, 100}

\bibitem[\protect\citeauthoryear{{Cutri} et~al.,}{{Cutri}
  et~al.}{2013}]{Cutri2013}
{Cutri} R.~M.,  et~al., 2013, Technical report, {Explanatory Supplement to the
  AllWISE Data Release Products}

\bibitem[\protect\citeauthoryear{{Danielson} et~al.,}{{Danielson}
  et~al.}{2017}]{Danielson2017}
{Danielson} A.~L.~R.,  et~al., 2017, \mndoi [\apj] {10.3847/1538-4357/aa6caf},
  \href {https://ui.adsabs.harvard.edu/abs/2017ApJ...840...78D} {840, 78}

\bibitem[\protect\citeauthoryear{{Dannerbauer}, {Lehnert}, {Lutz}, {Tacconi},
  {Bertoldi}, {Carilli}, {Genzel}  \& {Menten}}{{Dannerbauer}
  et~al.}{2002}]{Dannerbauer2002}
{Dannerbauer} H.,  {Lehnert} M.~D.,  {Lutz} D.,  {Tacconi} L.,  {Bertoldi} F.,
  {Carilli} C.,  {Genzel} R.,   {Menten} K.,  2002, \mndoi [\apj]
  {10.1086/340762}, \href
  {https://ui.adsabs.harvard.edu/abs/2002ApJ...573..473D} {573, 473}

\bibitem[\protect\citeauthoryear{{Dannerbauer}, {Lehnert}, {Lutz}, {Tacconi},
  {Bertoldi}, {Carilli}, {Genzel}  \& {Menten}}{{Dannerbauer}
  et~al.}{2004}]{Dannerbauer2004}
{Dannerbauer} H.,  {Lehnert} M.~D.,  {Lutz} D.,  {Tacconi} L.,  {Bertoldi} F.,
  {Carilli} C.,  {Genzel} R.,   {Menten} K.~M.,  2004, \mndoi [\apj]
  {10.1086/383138}, \href
  {https://ui.adsabs.harvard.edu/abs/2004ApJ...606..664D} {606, 664}

\bibitem[\protect\citeauthoryear{{Dannerbauer}, {Walter}  \&
  {Morrison}}{{Dannerbauer} et~al.}{2008}]{Dannerbauer2008}
{Dannerbauer} H.,  {Walter} F.,   {Morrison} G.,  2008, \mndoi [\apjl]
  {10.1086/528794}, \href
  {https://ui.adsabs.harvard.edu/abs/2008ApJ...673L.127D} {673, L127}

\bibitem[\protect\citeauthoryear{{Dannerbauer} et~al.,}{{Dannerbauer}
  et~al.}{2014}]{Dannerbauer2014}
{Dannerbauer} H.,  et~al., 2014, \mndoi [\aap] {10.1051/0004-6361/201423771},
  \href {https://ui.adsabs.harvard.edu/abs/2014A&A...570A..55D} {570, A55}

\bibitem[\protect\citeauthoryear{{Dannerbauer} et~al.,}{{Dannerbauer}
  et~al.}{2019a}]{Dannerbauer2019b}
{Dannerbauer} H.,  et~al., 2019a, BAAS, \href
  {https://ui.adsabs.harvard.edu/abs/2019BAAS...51c.293D} {51, 293}

\bibitem[\protect\citeauthoryear{{Dannerbauer}, {Harrington},
  {D{\'\i}az-S{\'a}nchez}, {Iglesias-Groth}, {Rebolo}, {Genova-Santos}  \&
  {Krips}}{{Dannerbauer} et~al.}{2019b}]{Dannerbauer2019a}
{Dannerbauer} H.,  {Harrington} K.,  {D{\'\i}az-S{\'a}nchez} A.,
  {Iglesias-Groth} S.,  {Rebolo} R.,  {Genova-Santos} R.~T.,   {Krips} M.,
  2019b, \mndoi [\aj] {10.3847/1538-3881/aaf50b}, \href
  {https://ui.adsabs.harvard.edu/abs/2019AJ....158...34D} {158, 34}

\bibitem[\protect\citeauthoryear{{De Looze} et~al.,}{{De Looze}
  et~al.}{2014}]{DeLooze2014}
{De Looze} I.,  et~al., 2014, \mndoi [\aap] {10.1051/0004-6361/201322489},
  \href {https://ui.adsabs.harvard.edu/abs/2014A&A...568A..62D} {568, A62}

\bibitem[\protect\citeauthoryear{{Delabrouille} et~al.,}{{Delabrouille}
  et~al.}{2019}]{Delabrouille2019}
{Delabrouille} J.,  et~al., 2019, arXiv e-prints, \href
  {https://ui.adsabs.harvard.edu/abs/2019arXiv190901591D} {p. arXiv:1909.01591}

\bibitem[\protect\citeauthoryear{{Dessauges-Zavadsky}
  et~al.,}{{Dessauges-Zavadsky} et~al.}{2019}]{Dessauges2019}
{Dessauges-Zavadsky} M.,  et~al., 2019, \mndoi [Nature Astronomy]
  {10.1038/s41550-019-0874-0}, \href
  {https://ui.adsabs.harvard.edu/abs/2019NatAs...3.1115D} {3, 1115}

\bibitem[\protect\citeauthoryear{{D{\'\i}az-S{\'a}nchez}, {Iglesias-Groth},
  {Rebolo}  \& {Dannerbauer}}{{D{\'\i}az-S{\'a}nchez}
  et~al.}{2017}]{DiazSanchez2017}
{D{\'\i}az-S{\'a}nchez} A.,  {Iglesias-Groth} S.,  {Rebolo} R.,   {Dannerbauer}
  H.,  2017, \mndoi [\apjl] {10.3847/2041-8213/aa79ef}, \href
  {https://ui.adsabs.harvard.edu/abs/2017ApJ...843L..22D} {843, L22}

\bibitem[\protect\citeauthoryear{{Dressler}}{{Dressler}}{1980}]{Dressler1980}
{Dressler} A.,  1980, \mndoi [\apj] {10.1086/157753}, \href
  {https://ui.adsabs.harvard.edu/abs/1980ApJ...236..351D} {236, 351}

\bibitem[\protect\citeauthoryear{{Dudzevi{\v{c}}i{\={u}}t{\.{e}}}
  et~al.,}{{Dudzevi{\v{c}}i{\={u}}t{\.{e}}} et~al.}{2020}]{Dudzeviciute2019}
{Dudzevi{\v{c}}i{\={u}}t{\.{e}}} U.,  et~al., 2020, \mndoi [\mnras]
  {10.1093/mnras/staa769}, \href
  {https://ui.adsabs.harvard.edu/abs/2020MNRAS.tmp.1074D} {}

\bibitem[\protect\citeauthoryear{{Dunlop} et~al.,}{{Dunlop}
  et~al.}{2004}]{Dunlop2004}
{Dunlop} J.~S.,  et~al., 2004, \mndoi [\mnras]
  {10.1111/j.1365-2966.2004.07700.x}, \href
  {https://ui.adsabs.harvard.edu/abs/2004MNRAS.350..769D} {350, 769}

\bibitem[\protect\citeauthoryear{{Eales} et~al.,}{{Eales}
  et~al.}{2010}]{Eales2010}
{Eales} S.,  et~al., 2010, \mndoi [\pasp] {10.1086/653086}, \href
  {https://ui.adsabs.harvard.edu/abs/2010PASP..122..499E} {122, 499}

\bibitem[\protect\citeauthoryear{{Flores-Cacho} et~al.,}{{Flores-Cacho}
  et~al.}{2016}]{FloresCacho2016}
{Flores-Cacho} I.,  et~al., 2016, \mndoi [\aap] {10.1051/0004-6361/201425226},
  \href {http://adsabs.harvard.edu/abs/2016A\%A...585A..54F} {585, A54}

\bibitem[\protect\citeauthoryear{{Franceschini}, {Toffolatti}, {Mazzei},
  {Danese}  \& {de Zotti}}{{Franceschini} et~al.}{1991}]{Franceschini1991}
{Franceschini} A.,  {Toffolatti} L.,  {Mazzei} P.,  {Danese} L.,   {de Zotti}
  G.,  1991, \aaps, \href
  {https://ui.adsabs.harvard.edu/abs/1991A&AS...89..285F} {89, 285}

\bibitem[\protect\citeauthoryear{{Fu} et~al.,}{{Fu} et~al.}{2012}]{Fu2012}
{Fu} H.,  et~al., 2012, \mndoi [\apj] {10.1088/0004-637X/753/2/134}, \href
  {https://ui.adsabs.harvard.edu/abs/2012ApJ...753..134F} {753, 134}

\bibitem[\protect\citeauthoryear{{Fudamoto} et~al.,}{{Fudamoto}
  et~al.}{2017}]{Fudamoto2017}
{Fudamoto} Y.,  et~al., 2017, \mndoi [\mnras] {10.1093/mnras/stx1956}, \href
  {https://ui.adsabs.harvard.edu/abs/2017MNRAS.472.2028F} {472, 2028}

\bibitem[\protect\citeauthoryear{{Fujimoto} et~al.,}{{Fujimoto}
  et~al.}{2019}]{Fujimoto2019}
{Fujimoto} S.,  et~al., 2019, \mndoi [\apj] {10.3847/1538-4357/ab480f}, \href
  {https://ui.adsabs.harvard.edu/abs/2019ApJ...887..107F} {887, 107}

\bibitem[\protect\citeauthoryear{{Gobat} et~al.,}{{Gobat}
  et~al.}{2011}]{Gobat2011}
{Gobat} R.,  et~al., 2011, \mndoi [\aap] {10.1051/0004-6361/201016084}, \href
  {https://ui.adsabs.harvard.edu/abs/2011A&A...526A.133G} {526, A133}

\bibitem[\protect\citeauthoryear{{G{\'o}mez-Guijarro}
  et~al.,}{{G{\'o}mez-Guijarro} et~al.}{2019}]{GomezGuijarro2019}
{G{\'o}mez-Guijarro} C.,  et~al., 2019, \mndoi [\apj]
  {10.3847/1538-4357/ab002a}, \href
  {http://adsabs.harvard.edu/abs/2019ApJ...872..117G} {872, 117}

\bibitem[\protect\citeauthoryear{{Gonzalez} et~al.,}{{Gonzalez}
  et~al.}{2019}]{Gonzalez2019}
{Gonzalez} A.~H.,  et~al., 2019, \mndoi [\apjs] {10.3847/1538-4365/aafad2},
  \href {https://ui.adsabs.harvard.edu/abs/2019ApJS..240...33G} {240, 33}

\bibitem[\protect\citeauthoryear{{Grandis}, {Mohr}, {Dietrich}, {Bocquet},
  {Saro}, {Klein}, {Paulus}  \& {Capasso}}{{Grandis}
  et~al.}{2019}]{Grandis2019}
{Grandis} S.,  {Mohr} J.~J.,  {Dietrich} J.~P.,  {Bocquet} S.,  {Saro} A.,
  {Klein} M.,  {Paulus} M.,   {Capasso} R.,  2019, \mndoi [\mnras]
  {10.1093/mnras/stz1778}, \href
  {https://ui.adsabs.harvard.edu/abs/2019MNRAS.488.2041G} {488, 2041}

\bibitem[\protect\citeauthoryear{{Gruppioni} et~al.,}{{Gruppioni}
  et~al.}{2013}]{Gruppioni2013}
{Gruppioni} C.,  et~al., 2013, \mndoi [\mnras] {10.1093/mnras/stt308}, \href
  {https://ui.adsabs.harvard.edu/abs/2013MNRAS.432...23G} {432, 23}

\bibitem[\protect\citeauthoryear{{Gruppioni} et~al.,}{{Gruppioni}
  et~al.}{2015}]{Gruppioni2015}
{Gruppioni} C.,  et~al., 2015, \mndoi [\mnras] {10.1093/mnras/stv1204}, \href
  {https://ui.adsabs.harvard.edu/abs/2015MNRAS.451.3419G} {451, 3419}

\bibitem[\protect\citeauthoryear{{Gullberg} et~al.,}{{Gullberg}
  et~al.}{2015}]{Gullberg2015}
{Gullberg} B.,  et~al., 2015, \mndoi [\mnras] {10.1093/mnras/stv372}, \href
  {https://ui.adsabs.harvard.edu/abs/2015MNRAS.449.2883G} {449, 2883}

\bibitem[\protect\citeauthoryear{{Hanany} et~al.,}{{Hanany}
  et~al.}{2019}]{Hanany2019}
{Hanany} S.,  et~al., 2019, arXiv e-prints, \href
  {https://ui.adsabs.harvard.edu/abs/2019arXiv190210541H} {p. arXiv:1902.10541}

\bibitem[\protect\citeauthoryear{{Harrington} et~al.,}{{Harrington}
  et~al.}{2016}]{Harrington2016}
{Harrington} K.~C.,  et~al., 2016, \mndoi [\mnras] {10.1093/mnras/stw614},
  \href {https://ui.adsabs.harvard.edu/abs/2016MNRAS.458.4383H} {458, 4383}

\bibitem[\protect\citeauthoryear{{Harrington} et~al.,}{{Harrington}
  et~al.}{2018}]{Harrington2018}
{Harrington} K.~C.,  et~al., 2018, \mndoi [\mnras] {10.1093/mnras/stx3043},
  \href {https://ui.adsabs.harvard.edu/abs/2018MNRAS.474.3866H} {474, 3866}

\bibitem[\protect\citeauthoryear{{Hashimoto} et~al.,}{{Hashimoto}
  et~al.}{2018}]{Hashimoto2018}
{Hashimoto} T.,  et~al., 2018, \mndoi [\nat] {10.1038/s41586-018-0117-z}, \href
  {https://ui.adsabs.harvard.edu/abs/2018Natur.557..392H} {557, 392}

\bibitem[\protect\citeauthoryear{{Heckman} \& {Best}}{{Heckman} \&
  {Best}}{2014}]{Heckman2014}
{Heckman} T.~M.,  {Best} P.~N.,  2014, \mndoi [\araa]
  {10.1146/annurev-astro-081913-035722}, \href
  {https://ui.adsabs.harvard.edu/abs/2014ARA&A..52..589H} {52, 589}

\bibitem[\protect\citeauthoryear{{Hezaveh} et~al.,}{{Hezaveh}
  et~al.}{2013}]{Hezaveh2013}
{Hezaveh} Y.~D.,  et~al., 2013, \mndoi [\apj] {10.1088/0004-637X/767/2/132},
  \href {https://ui.adsabs.harvard.edu/abs/2013ApJ...767..132H} {767, 132}

\bibitem[\protect\citeauthoryear{{Higuchi} et~al.,}{{Higuchi}
  et~al.}{2019}]{Higuchi2019}
{Higuchi} R.,  et~al., 2019, \mndoi [\apj] {10.3847/1538-4357/ab2192}, \href
  {https://ui.adsabs.harvard.edu/abs/2019ApJ...879...28H} {879, 28}

\bibitem[\protect\citeauthoryear{{Hill} et~al.,}{{Hill}
  et~al.}{2020}]{Hill2020}
{Hill} R.,  et~al., 2020, arXiv e-prints, \href
  {https://ui.adsabs.harvard.edu/abs/2020arXiv200211600H} {p. arXiv:2002.11600}

\bibitem[\protect\citeauthoryear{{Hilton} et~al.,}{{Hilton}
  et~al.}{2018}]{Hilton2018}
{Hilton} M.,  et~al., 2018, \mndoi [\apjs] {10.3847/1538-4365/aaa6cb}, \href
  {https://ui.adsabs.harvard.edu/abs/2018ApJS..235...20H} {235, 20}

\bibitem[\protect\citeauthoryear{{Hughes} et~al.,}{{Hughes}
  et~al.}{1998}]{Hughes1998}
{Hughes} D.~H.,  et~al., 1998, \mndoi [\nat] {10.1038/28328}, \href
  {https://ui.adsabs.harvard.edu/abs/1998Natur.394..241H} {394, 241}

\bibitem[\protect\citeauthoryear{{Hung} et~al.,}{{Hung}
  et~al.}{2016}]{Hung2016}
{Hung} C.-L.,  et~al., 2016, \mndoi [\apj] {10.3847/0004-637X/826/2/130}, \href
  {https://ui.adsabs.harvard.edu/abs/2016ApJ...826..130H} {826, 130}

\bibitem[\protect\citeauthoryear{{Inoue} et~al.,}{{Inoue}
  et~al.}{2016}]{Inoue2016}
{Inoue} A.~K.,  et~al., 2016, \mndoi [Science] {10.1126/science.aaf0714}, \href
  {https://ui.adsabs.harvard.edu/abs/2016Sci...352.1559I} {352, 1559}

\bibitem[\protect\citeauthoryear{{Ivison}, {Smail}, {Le Borgne}, {Blain},
  {Kneib}, {Bezecourt}, {Kerr}  \& {Davies}}{{Ivison}
  et~al.}{1998}]{Ivison1998}
{Ivison} R.~J.,  {Smail} I.,  {Le Borgne} J.~F.,  {Blain} A.~W.,  {Kneib}
  J.~P.,  {Bezecourt} J.,  {Kerr} T.~H.,   {Davies} J.~K.,  1998, \mndoi
  [\mnras] {10.1046/j.1365-8711.1998.01677.x}, \href
  {https://ui.adsabs.harvard.edu/abs/1998MNRAS.298..583I} {298, 583}

\bibitem[\protect\citeauthoryear{{Ivison} et~al.,}{{Ivison}
  et~al.}{2002}]{Ivison2002}
{Ivison} R.~J.,  et~al., 2002, \mndoi [\mnras]
  {10.1046/j.1365-8711.2002.05900.x}, \href
  {https://ui.adsabs.harvard.edu/abs/2002MNRAS.337....1I} {337, 1}

\bibitem[\protect\citeauthoryear{{Ivison} et~al.,}{{Ivison}
  et~al.}{2013}]{Ivison2013}
{Ivison} R.~J.,  et~al., 2013, \mndoi [\apj] {10.1088/0004-637X/772/2/137},
  \href {http://adsabs.harvard.edu/abs/2013ApJ...772..137I} {772, 137}

\bibitem[\protect\citeauthoryear{{Ivison}, {Page}, {Cirasuolo}, {Harrison},
  {Mainieri}, {Arumugam}  \& {Dudzevi{\v{c}}i{\={u}}t{\.{e}}}}{{Ivison}
  et~al.}{2019}]{Ivison2019}
{Ivison} R.~J.,  {Page} M.~J.,  {Cirasuolo} M.,  {Harrison} C.~M.,  {Mainieri}
  V.,  {Arumugam} V.,   {Dudzevi{\v{c}}i{\={u}}t{\.{e}}} U.,  2019, \mndoi
  [\mnras] {10.1093/mnras/stz2180}, \href
  {https://ui.adsabs.harvard.edu/abs/2019MNRAS.tmp.2100I} {p.~2100}

\bibitem[\protect\citeauthoryear{{Jones}, {Maiolino}, {Caselli}  \&
  {Carniani}}{{Jones} et~al.}{2019}]{Jones2019}
{Jones} G.~C.,  {Maiolino} R.,  {Caselli} P.,   {Carniani} S.,  2019, \mndoi
  [\aap] {10.1051/0004-6361/201936989}, \href
  {https://ui.adsabs.harvard.edu/abs/2019A&A...632L...7J} {632, L7}

\bibitem[\protect\citeauthoryear{{Kato} et~al.,}{{Kato}
  et~al.}{2016}]{Kato2016}
{Kato} Y.,  et~al., 2016, \mndoi [\mnras] {10.1093/mnras/stw1237}, \href
  {https://ui.adsabs.harvard.edu/abs/2016MNRAS.460.3861K} {460, 3861}

\bibitem[\protect\citeauthoryear{{Kauffmann}, {White}, {Heckman}, {M{\'e}nard},
  {Brinchmann}, {Charlot}, {Tremonti}  \& {Brinkmann}}{{Kauffmann}
  et~al.}{2004}]{Kauffmann2004}
{Kauffmann} G.,  {White} S. D.~M.,  {Heckman} T.~M.,  {M{\'e}nard} B.,
  {Brinchmann} J.,  {Charlot} S.,  {Tremonti} C.,   {Brinkmann} J.,  2004,
  \mndoi [\mnras] {10.1111/j.1365-2966.2004.08117.x}, \href
  {https://ui.adsabs.harvard.edu/abs/2004MNRAS.353..713K} {353, 713}

\bibitem[\protect\citeauthoryear{{Kaviani}, {Haehnelt}  \&
  {Kauffmann}}{{Kaviani} et~al.}{2003}]{Kaviani2003}
{Kaviani} A.,  {Haehnelt} M.~G.,   {Kauffmann} G.,  2003, \mndoi [\mnras]
  {10.1046/j.1365-8711.2003.06318.x}, \href
  {https://ui.adsabs.harvard.edu/abs/2003MNRAS.340..739K} {340, 739}

\bibitem[\protect\citeauthoryear{{Kennicutt} \& {Evans}}{{Kennicutt} \&
  {Evans}}{2012}]{KennicuttEvans2012}
{Kennicutt} R.~C.,  {Evans} N.~J.,  2012, \mndoi [\araa]
  {10.1146/annurev-astro-081811-125610}, \href
  {https://ui.adsabs.harvard.edu/abs/2012ARA&A..50..531K} {50, 531}

\bibitem[\protect\citeauthoryear{{King} \& {Pounds}}{{King} \&
  {Pounds}}{2015}]{KingPounds2015}
{King} A.,  {Pounds} K.,  2015, \mndoi [\araa]
  {10.1146/annurev-astro-082214-122316}, \href
  {http://adsabs.harvard.edu/abs/2015ARA\%A..53..115K} {53, 115}

\bibitem[\protect\citeauthoryear{{Klein} et~al.,}{{Klein}
  et~al.}{2019}]{Klein2019}
{Klein} M.,  et~al., 2019, \mndoi [\mnras] {10.1093/mnras/stz1463}, \href
  {https://ui.adsabs.harvard.edu/abs/2019MNRAS.488..739K} {488, 739}

\bibitem[\protect\citeauthoryear{{Kneissl} et~al.,}{{Kneissl}
  et~al.}{2019}]{Kneissl2019}
{Kneissl} R.,  et~al., 2019, \mndoi [\aap] {10.1051/0004-6361/201833252}, \href
  {https://ui.adsabs.harvard.edu/abs/2019A&A...625A..96K} {625, A96}

\bibitem[\protect\citeauthoryear{{Kravtsov} \& {Borgani}}{{Kravtsov} \&
  {Borgani}}{2012}]{KravtsovBorgani2012}
{Kravtsov} A.~V.,  {Borgani} S.,  2012, \mndoi [\araa]
  {10.1146/annurev-astro-081811-125502}, \href
  {https://ui.adsabs.harvard.edu/abs/2012ARA&A..50..353K} {50, 353}

\bibitem[\protect\citeauthoryear{{Kubo} et~al.,}{{Kubo}
  et~al.}{2019}]{Kubo2019}
{Kubo} M.,  et~al., 2019, \mndoi [\apj] {10.3847/1538-4357/ab5a80}, \href
  {https://ui.adsabs.harvard.edu/abs/2019ApJ...887..214K} {887, 214}

\bibitem[\protect\citeauthoryear{{LSST Science Collaboration}}{{LSST Science
  Collaboration}}{2009}]{LSST2009}
{LSST Science Collaboration} 2009, arXiv e-prints, \href
  {https://ui.adsabs.harvard.edu/abs/2009arXiv0912.0201L} {p. arXiv:0912.0201}

\bibitem[\protect\citeauthoryear{{Lacaille} et~al.,}{{Lacaille}
  et~al.}{2019}]{Lacaille2018}
{Lacaille} K.~M.,  et~al., 2019, \mndoi [\mnras] {10.1093/mnras/stz1742}, \href
  {https://ui.adsabs.harvard.edu/abs/2019MNRAS.488.1790L} {488, 1790}

\bibitem[\protect\citeauthoryear{{Laporte} et~al.,}{{Laporte}
  et~al.}{2017}]{Laporte2017}
{Laporte} N.,  et~al., 2017, \mndoi [\apjl] {10.3847/2041-8213/aa62aa}, \href
  {https://ui.adsabs.harvard.edu/abs/2017ApJ...837L..21L} {837, L21}

\bibitem[\protect\citeauthoryear{{Laureijs} et~al.,}{{Laureijs}
  et~al.}{2011}]{Laureijs2011}
{Laureijs} R.,  et~al., 2011, arXiv e-prints, \href
  {https://ui.adsabs.harvard.edu/abs/2011arXiv1110.3193L} {p. arXiv:1110.3193}

\bibitem[\protect\citeauthoryear{{Lemaux} et~al.,}{{Lemaux}
  et~al.}{2019}]{Lemaux2019}
{Lemaux} B.~C.,  et~al., 2019, \mndoi [\mnras] {10.1093/mnras/stz2661}, \href
  {https://ui.adsabs.harvard.edu/abs/2019MNRAS.490.1231L} {490, 1231}

\bibitem[\protect\citeauthoryear{{Lima}, {Jain}  \& {Devlin}}{{Lima}
  et~al.}{2010}]{Lima2010}
{Lima} M.,  {Jain} B.,   {Devlin} M.,  2010, \mndoi [\mnras]
  {10.1111/j.1365-2966.2010.16884.x}, \href
  {https://ui.adsabs.harvard.edu/abs/2010MNRAS.406.2352L} {406, 2352}

\bibitem[\protect\citeauthoryear{{Lupu} et~al.,}{{Lupu}
  et~al.}{2012}]{Lupu2012}
{Lupu} R.~E.,  et~al., 2012, \mndoi [\apj] {10.1088/0004-637X/757/2/135}, \href
  {https://ui.adsabs.harvard.edu/abs/2012ApJ...757..135L} {757, 135}

\bibitem[\protect\citeauthoryear{{Lutz}}{{Lutz}}{2014}]{Lutz2014}
{Lutz} D.,  2014, \mndoi [\araa] {10.1146/annurev-astro-081913-035953}, \href
  {https://ui.adsabs.harvard.edu/abs/2014ARA&A..52..373L} {52, 373}

\bibitem[\protect\citeauthoryear{{Magnelli} et~al.,}{{Magnelli}
  et~al.}{2013}]{Magnelli2013}
{Magnelli} B.,  et~al., 2013, \mndoi [\aap] {10.1051/0004-6361/201321371},
  \href {https://ui.adsabs.harvard.edu/abs/2013A&A...553A.132M} {553, A132}

\bibitem[\protect\citeauthoryear{{Mancuso} et~al.,}{{Mancuso}
  et~al.}{2015}]{Mancuso2015}
{Mancuso} C.,  et~al., 2015, \mndoi [\apj] {10.1088/0004-637X/810/1/72}, \href
  {https://ui.adsabs.harvard.edu/abs/2015ApJ...810...72M} {810, 72}

\bibitem[\protect\citeauthoryear{{Massardi} et~al.,}{{Massardi}
  et~al.}{2018}]{Massardi2018}
{Massardi} M.,  et~al., 2018, \mndoi [\aap] {10.1051/0004-6361/201731751},
  \href {https://ui.adsabs.harvard.edu/abs/2018A&A...610A..53M} {610, A53}

\bibitem[\protect\citeauthoryear{{Meixner} et~al.,}{{Meixner}
  et~al.}{2019}]{Cooray2019}
{Meixner} M.,  et~al., 2019, arXiv e-prints, \href
  {https://ui.adsabs.harvard.edu/abs/2019arXiv191206213M} {p. arXiv:1912.06213}

\bibitem[\protect\citeauthoryear{{Miller} et~al.,}{{Miller}
  et~al.}{2018}]{Miller2018}
{Miller} T.~B.,  et~al., 2018, \mndoi [\nat] {10.1038/s41586-018-0025-2}, \href
  {http://adsabs.harvard.edu/abs/2018Natur.556..469M} {556, 469}

\bibitem[\protect\citeauthoryear{{Mocanu} et~al.,}{{Mocanu}
  et~al.}{2013}]{Mocanu2013}
{Mocanu} L.~M.,  et~al., 2013, \mndoi [\apj] {10.1088/0004-637X/779/1/61},
  \href {https://ui.adsabs.harvard.edu/abs/2013ApJ...779...61M} {779, 61}

\bibitem[\protect\citeauthoryear{{Nantais} et~al.,}{{Nantais}
  et~al.}{2017}]{Nantais2017}
{Nantais} J.~B.,  et~al., 2017, \mndoi [\mnras] {10.1093/mnrasl/slw224}, \href
  {https://ui.adsabs.harvard.edu/abs/2017MNRAS.465L.104N} {465, L104}

\bibitem[\protect\citeauthoryear{{Nayyeri} et~al.,}{{Nayyeri}
  et~al.}{2016}]{Nayyeri2016}
{Nayyeri} H.,  et~al., 2016, \mndoi [\apj] {10.3847/0004-637X/823/1/17}, \href
  {https://ui.adsabs.harvard.edu/abs/2016ApJ...823...17N} {823, 17}

\bibitem[\protect\citeauthoryear{{Negrello}, {Gonz{\'a}lez-Nuevo},
  {Magliocchetti}, {Moscardini}, {De Zotti}, {Toffolatti}  \&
  {Danese}}{{Negrello} et~al.}{2005}]{Negrello2005}
{Negrello} M.,  {Gonz{\'a}lez-Nuevo} J.,  {Magliocchetti} M.,  {Moscardini} L.,
   {De Zotti} G.,  {Toffolatti} L.,   {Danese} L.,  2005, \mndoi [\mnras]
  {10.1111/j.1365-2966.2005.08783.x}, \href
  {https://ui.adsabs.harvard.edu/abs/2005MNRAS.358..869N} {358, 869}

\bibitem[\protect\citeauthoryear{{Negrello}, {Perrotta}, {Gonz{\'a}lez-Nuevo},
  {Silva}, {de Zotti}, {Granato}, {Baccigalupi}  \& {Danese}}{{Negrello}
  et~al.}{2007}]{Negrello2007}
{Negrello} M.,  {Perrotta} F.,  {Gonz{\'a}lez-Nuevo} J.,  {Silva} L.,  {de
  Zotti} G.,  {Granato} G.~L.,  {Baccigalupi} C.,   {Danese} L.,  2007, \mndoi
  [\mnras] {10.1111/j.1365-2966.2007.11708.x}, \href
  {https://ui.adsabs.harvard.edu/abs/2007MNRAS.377.1557N} {377, 1557}

\bibitem[\protect\citeauthoryear{{Negrello} et~al.,}{{Negrello}
  et~al.}{2010}]{Negrello2010}
{Negrello} M.,  et~al., 2010, \mndoi [Science] {10.1126/science.1193420}, \href
  {https://ui.adsabs.harvard.edu/abs/2010Sci...330..800N} {330, 800}

\bibitem[\protect\citeauthoryear{{Negrello} et~al.,}{{Negrello}
  et~al.}{2014}]{Negrello2014}
{Negrello} M.,  et~al., 2014, \mndoi [\mnras] {10.1093/mnras/stu413}, \href
  {https://ui.adsabs.harvard.edu/abs/2014MNRAS.440.1999N} {440, 1999}

\bibitem[\protect\citeauthoryear{{Negrello} et~al.,}{{Negrello}
  et~al.}{2017a}]{Negrello2017lensed}
{Negrello} M.,  et~al., 2017a, \mndoi [\mnras] {10.1093/mnras/stw2911}, \href
  {https://ui.adsabs.harvard.edu/abs/2017MNRAS.465.3558N} {465, 3558}

\bibitem[\protect\citeauthoryear{{Negrello} et~al.,}{{Negrello}
  et~al.}{2017b}]{Negrello2017protocl}
{Negrello} M.,  et~al., 2017b, \mndoi [\mnras] {10.1093/mnras/stx1367}, \href
  {https://ui.adsabs.harvard.edu/abs/2017MNRAS.470.2253N} {470, 2253}

\bibitem[\protect\citeauthoryear{{Neri} et~al.,}{{Neri}
  et~al.}{2020}]{Neri2019}
{Neri} R.,  et~al., 2020, \mndoi [\aap] {10.1051/0004-6361/201936988}, \href
  {https://ui.adsabs.harvard.edu/abs/2020A&A...635A...7N} {635, A7}

\bibitem[\protect\citeauthoryear{{Nesvadba}, {Ca{\~n}ameras}, {Kneissl},
  {Koenig}, {Yang}, {Le Floc'h}, {Omont}  \& {Scott}}{{Nesvadba}
  et~al.}{2019}]{Nesvadba2019}
{Nesvadba} N.~P.~H.,  {Ca{\~n}ameras} R.,  {Kneissl} R.,  {Koenig} S.,  {Yang}
  C.,  {Le Floc'h} E.,  {Omont} A.,   {Scott} D.,  2019, \mndoi [\aap]
  {10.1051/0004-6361/201833777}, \href
  {https://ui.adsabs.harvard.edu/abs/2019A&A...624A..23N} {624, A23}

\bibitem[\protect\citeauthoryear{{Nguyen} et~al.,}{{Nguyen}
  et~al.}{2010}]{Nguyen2010}
{Nguyen} H.~T.,  et~al., 2010, \mndoi [\aap] {10.1051/0004-6361/201014680},
  \href {https://ui.adsabs.harvard.edu/abs/2010A&A...518L...5N} {518, L5}

\bibitem[\protect\citeauthoryear{{Niemi}, {Somerville}, {Ferguson}, {Huang},
  {Lotz}  \& {Koekemoer}}{{Niemi} et~al.}{2012}]{Niemi2012}
{Niemi} S.-M.,  {Somerville} R.~S.,  {Ferguson} H.~C.,  {Huang} K.-H.,  {Lotz}
  J.,   {Koekemoer} A.~M.,  2012, \mndoi [\mnras]
  {10.1111/j.1365-2966.2012.20425.x}, \href
  {https://ui.adsabs.harvard.edu/abs/2012MNRAS.421.1539N} {421, 1539}

\bibitem[\protect\citeauthoryear{{Oguri} et~al.,}{{Oguri}
  et~al.}{2018}]{Oguri2018}
{Oguri} M.,  et~al., 2018, \mndoi [\pasj] {10.1093/pasj/psx042}, \href
  {https://ui.adsabs.harvard.edu/abs/2018PASJ...70S..20O} {70, S20}

\bibitem[\protect\citeauthoryear{{Oliver} et~al.,}{{Oliver}
  et~al.}{2012}]{Oliver2012}
{Oliver} S.~J.,  et~al., 2012, \mndoi [\mnras]
  {10.1111/j.1365-2966.2012.20912.x}, \href
  {https://ui.adsabs.harvard.edu/abs/2012MNRAS.424.1614O} {424, 1614}

\bibitem[\protect\citeauthoryear{{Oteo} et~al.,}{{Oteo}
  et~al.}{2018}]{Oteo2018}
{Oteo} I.,  et~al., 2018, \mndoi [\apj] {10.3847/1538-4357/aaa1f1}, \href
  {http://adsabs.harvard.edu/abs/2018ApJ...856...72O} {856, 72}

\bibitem[\protect\citeauthoryear{{Overzier}}{{Overzier}}{2016}]{Overzier2016}
{Overzier} R.~A.,  2016, \mndoi [\aapr] {10.1007/s00159-016-0100-3}, \href
  {http://adsabs.harvard.edu/abs/2016A\%ARv..24...14O} {24, 14}

\bibitem[\protect\citeauthoryear{{Overzier} \& {Kashikawa}}{{Overzier} \&
  {Kashikawa}}{2019}]{OverzierKashikawa2019}
{Overzier} R.,  {Kashikawa} N.,  2019, \baas, \href
  {https://ui.adsabs.harvard.edu/abs/2019BAAS...51c.180O} {51, 180}

\bibitem[\protect\citeauthoryear{{Paciga}, {Scott}  \& {Chapin}}{{Paciga}
  et~al.}{2009}]{Paciga2009}
{Paciga} G.,  {Scott} D.,   {Chapin} E.~L.,  2009, \mndoi [\mnras]
  {10.1111/j.1365-2966.2009.14627.x}, \href
  {https://ui.adsabs.harvard.edu/abs/2009MNRAS.395.1153P} {395, 1153}

\bibitem[\protect\citeauthoryear{{Papadopoulos}, {Thi}, {Miniati}  \&
  {Viti}}{{Papadopoulos} et~al.}{2011}]{Papadopoulos2011}
{Papadopoulos} P.~P.,  {Thi} W.-F.,  {Miniati} F.,   {Viti} S.,  2011, \mndoi
  [\mnras] {10.1111/j.1365-2966.2011.18504.x}, \href
  {https://ui.adsabs.harvard.edu/abs/2011MNRAS.414.1705P} {414, 1705}

\bibitem[\protect\citeauthoryear{{Perrotta}, {Baccigalupi}, {Bartelmann}, {De
  Zotti}  \& {Granato}}{{Perrotta} et~al.}{2002}]{Perotta2002}
{Perrotta} F.,  {Baccigalupi} C.,  {Bartelmann} M.,  {De Zotti} G.,   {Granato}
  G.~L.,  2002, \mndoi [\mnras] {10.1046/j.1365-8711.2002.05009.x}, \href
  {https://ui.adsabs.harvard.edu/abs/2002MNRAS.329..445P} {329, 445}

\bibitem[\protect\citeauthoryear{{Planck Collaboration Int. XXVII}}{{Planck
  Collaboration Int. XXVII}}{2015}]{PlanckHerschel2015}
{Planck Collaboration Int. XXVII} 2015, \mndoi [\aap]
  {10.1051/0004-6361/201424790}, \href
  {https://ui.adsabs.harvard.edu/abs/2015A&A...582A..30P} {582, A30}

\bibitem[\protect\citeauthoryear{{Planck Collaboration Int. XXXIX}}{{Planck
  Collaboration Int. XXXIX}}{2016}]{Planck2016highz}
{Planck Collaboration Int. XXXIX} 2016, \mndoi [\aap]
  {10.1051/0004-6361/201527206}, \href
  {https://ui.adsabs.harvard.edu/abs/2016A&A...596A.100P} {596, A100}

\bibitem[\protect\citeauthoryear{{Planck Collaboration VI}}{{Planck
  Collaboration VI}}{2018}]{PlanckCollaboration2018parameters}
{Planck Collaboration VI} 2018, arXiv e-prints, \href
  {https://ui.adsabs.harvard.edu/abs/2018arXiv180706209P} {p. arXiv:1807.06209}

\bibitem[\protect\citeauthoryear{{Planck Collaboration XXIV}}{{Planck
  Collaboration XXIV}}{2016}]{PlanckCollaboration2016SZcosm}
{Planck Collaboration XXIV} 2016, \mndoi [\aap] {10.1051/0004-6361/201525833},
  \href {https://ui.adsabs.harvard.edu/abs/2016A&A...594A..24P} {594, A24}

\bibitem[\protect\citeauthoryear{{Planck Collaboration XXVII}}{{Planck
  Collaboration XXVII}}{2016}]{PlanckCollaboration2016SZ}
{Planck Collaboration XXVII} 2016, \mndoi [\aap] {10.1051/0004-6361/201525823},
  \href {https://ui.adsabs.harvard.edu/abs/2016A&A...594A..27P} {594, A27}

\bibitem[\protect\citeauthoryear{{Pope}, {Borys}, {Scott}, {Conselice},
  {Dickinson}  \& {Mobasher}}{{Pope} et~al.}{2005}]{Pope2005}
{Pope} A.,  {Borys} C.,  {Scott} D.,  {Conselice} C.,  {Dickinson} M.,
  {Mobasher} B.,  2005, \mndoi [\mnras] {10.1111/j.1365-2966.2005.08759.x},
  \href {https://ui.adsabs.harvard.edu/abs/2005MNRAS.358..149P} {358, 149}

\bibitem[\protect\citeauthoryear{{Pope} et~al.,}{{Pope}
  et~al.}{2006}]{Pope2006}
{Pope} A.,  et~al., 2006, \mndoi [\mnras] {10.1111/j.1365-2966.2006.10575.x},
  \href {https://ui.adsabs.harvard.edu/abs/2006MNRAS.370.1185P} {370, 1185}

\bibitem[\protect\citeauthoryear{{Prandoni} \& {Seymour}}{{Prandoni} \&
  {Seymour}}{2015}]{PrandoniSeymour2015}
{Prandoni} I.,  {Seymour} N.,  2015, in Advancing Astrophysics with the Square
  Kilometre Array (AASKA14). p.~67 (\mn@eprint {arXiv} {1412.6512})

\bibitem[\protect\citeauthoryear{{Riechers} et~al.,}{{Riechers}
  et~al.}{2013}]{Riechers2013}
{Riechers} D.~A.,  et~al., 2013, \mndoi [\nat] {10.1038/nature12050}, \href
  {https://ui.adsabs.harvard.edu/abs/2013Natur.496..329R} {496, 329}

\bibitem[\protect\citeauthoryear{{Rodr{\'\i}guez-Mu{\~n}oz}
  et~al.,}{{Rodr{\'\i}guez-Mu{\~n}oz} et~al.}{2019}]{RodriguezMunoz2019}
{Rodr{\'\i}guez-Mu{\~n}oz} L.,  et~al., 2019, \mndoi [\mnras]
  {10.1093/mnras/sty3335}, \href
  {https://ui.adsabs.harvard.edu/abs/2019MNRAS.485..586R} {485, 586}

\bibitem[\protect\citeauthoryear{{Sartoris} et~al.,}{{Sartoris}
  et~al.}{2016}]{Sartoris2016}
{Sartoris} B.,  et~al., 2016, \mndoi [\mnras] {10.1093/mnras/stw630}, \href
  {https://ui.adsabs.harvard.edu/abs/2016MNRAS.459.1764S} {459, 1764}

\bibitem[\protect\citeauthoryear{{Schaerer} et~al.,}{{Schaerer}
  et~al.}{2020}]{Schaerer2020}
{Schaerer} D.,  et~al., 2020, arXiv e-prints, \href
  {https://ui.adsabs.harvard.edu/abs/2020arXiv200200979S} {p. arXiv:2002.00979}

\bibitem[\protect\citeauthoryear{{Sheth} \& {Tormen}}{{Sheth} \&
  {Tormen}}{1999}]{ShethTormen1999}
{Sheth} R.~K.,  {Tormen} G.,  1999, \mndoi [\mnras]
  {10.1046/j.1365-8711.1999.02692.x}, \href
  {https://ui.adsabs.harvard.edu/abs/1999MNRAS.308..119S} {308, 119}

\bibitem[\protect\citeauthoryear{{Sheth}, {Mo}  \& {Tormen}}{{Sheth}
  et~al.}{2001}]{Sheth2001}
{Sheth} R.~K.,  {Mo} H.~J.,   {Tormen} G.,  2001, \mndoi [\mnras]
  {10.1046/j.1365-8711.2001.04006.x}, \href
  {https://ui.adsabs.harvard.edu/abs/2001MNRAS.323....1S} {323, 1}

\bibitem[\protect\citeauthoryear{{Silva} et~al.,}{{Silva}
  et~al.}{2019}]{Silva2019}
{Silva} M.~B.,  et~al., 2019, arXiv e-prints, \href
  {https://ui.adsabs.harvard.edu/abs/2019arXiv190807533S} {p. arXiv:1908.07533}

\bibitem[\protect\citeauthoryear{{Smail}, {Ivison}  \& {Blain}}{{Smail}
  et~al.}{1997}]{Smail1997}
{Smail} I.,  {Ivison} R.~J.,   {Blain} A.~W.,  1997, \mndoi [\apjl]
  {10.1086/311017}, \href
  {https://ui.adsabs.harvard.edu/abs/1997ApJ...490L...5S} {490, L5}

\bibitem[\protect\citeauthoryear{{Smit} et~al.,}{{Smit}
  et~al.}{2018}]{Smit2018}
{Smit} R.,  et~al., 2018, \mndoi [\nat] {10.1038/nature24631}, \href
  {https://ui.adsabs.harvard.edu/abs/2018Natur.553..178S} {553, 178}

\bibitem[\protect\citeauthoryear{{Somerville}, {Gilmore}, {Primack}  \&
  {Dom{\'\i}nguez}}{{Somerville} et~al.}{2012}]{Somerville2012}
{Somerville} R.~S.,  {Gilmore} R.~C.,  {Primack} J.~R.,   {Dom{\'\i}nguez} A.,
  2012, \mndoi [\mnras] {10.1111/j.1365-2966.2012.20490.x}, \href
  {https://ui.adsabs.harvard.edu/abs/2012MNRAS.423.1992S} {423, 1992}

\bibitem[\protect\citeauthoryear{{Spergel} et~al.,}{{Spergel}
  et~al.}{2015}]{Spergel2015}
{Spergel} D.,  et~al., 2015, arXiv e-prints, \href
  {https://ui.adsabs.harvard.edu/abs/2015arXiv150303757S} {p. arXiv:1503.03757}

\bibitem[\protect\citeauthoryear{{Spilker}, {Aravena}  \&
  {B{\'e}thermin}}{{Spilker} et~al.}{2018}]{Spilker2018}
{Spilker} J.~S.,  {Aravena} M.,   {B{\'e}thermin} M. e.~a.,  2018, \mndoi
  [Science] {10.1126/science.aap8900}, \href
  {http://adsabs.harvard.edu/abs/2018Sci...361.1016S} {361, 1016}

\bibitem[\protect\citeauthoryear{{Springel} et~al.,}{{Springel}
  et~al.}{2005}]{Springel2005}
{Springel} V.,  et~al., 2005, \mndoi [\nat] {10.1038/nature03597}, \href
  {https://ui.adsabs.harvard.edu/abs/2005Natur.435..629S} {435, 629}

\bibitem[\protect\citeauthoryear{{Sridhar}, {Maurogordato}, {Benoist}, {Cappi}
  \& {Marulli}}{{Sridhar} et~al.}{2017}]{Sridhar2017}
{Sridhar} S.,  {Maurogordato} S.,  {Benoist} C.,  {Cappi} A.,   {Marulli} F.,
  2017, \mndoi [\aap] {10.1051/0004-6361/201629369}, \href
  {https://ui.adsabs.harvard.edu/abs/2017A&A...600A..32S} {600, A32}

\bibitem[\protect\citeauthoryear{{Stacey}, {Hailey-Dunsheath}, {Ferkinhoff},
  {Nikola}, {Parshley}, {Benford}, {Staguhn}  \& {Fiolet}}{{Stacey}
  et~al.}{2010}]{Stacey2010}
{Stacey} G.~J.,  {Hailey-Dunsheath} S.,  {Ferkinhoff} C.,  {Nikola} T.,
  {Parshley} S.~C.,  {Benford} D.~J.,  {Staguhn} J.~G.,   {Fiolet} N.,  2010,
  \mndoi [\apj] {10.1088/0004-637X/724/2/957}, \href
  {https://ui.adsabs.harvard.edu/abs/2010ApJ...724..957S} {724, 957}

\bibitem[\protect\citeauthoryear{{Sutter}}{{Sutter}}{2019}]{Sutter2019}
{Sutter} J.,  2019, in Linking Galaxies from the Epoch of Initial Star
  Formation to Today. p.~4 (\mn@eprint {arXiv} {1910.05416}),
  \mndoi{10.5281/zenodo.2635203}

\bibitem[\protect\citeauthoryear{{Swinbank}, {Smail}, {Chapman}, {Blain},
  {Ivison}  \& {Keel}}{{Swinbank} et~al.}{2004}]{Swinbank2004}
{Swinbank} A.~M.,  {Smail} I.,  {Chapman} S.~C.,  {Blain} A.~W.,  {Ivison}
  R.~J.,   {Keel} W.~C.,  2004, \mndoi [\apj] {10.1086/425171}, \href
  {https://ui.adsabs.harvard.edu/abs/2004ApJ...617...64S} {617, 64}

\bibitem[\protect\citeauthoryear{{Symeonidis}}{{Symeonidis}}{2017}]{Symeonidis2017}
{Symeonidis} M.,  2017, \mndoi [\mnras] {10.1093/mnras/stw2784}, \href
  {https://ui.adsabs.harvard.edu/abs/2017MNRAS.465.1401S} {465, 1401}

\bibitem[\protect\citeauthoryear{{Symeonidis}, {Giblin}, {Page}, {Pearson},
  {Bendo}, {Seymour}  \& {Oliver}}{{Symeonidis} et~al.}{2016}]{Symeonidis2016}
{Symeonidis} M.,  {Giblin} B.~M.,  {Page} M.~J.,  {Pearson} C.,  {Bendo} G.,
  {Seymour} N.,   {Oliver} S.~J.,  2016, \mndoi [\mnras]
  {10.1093/mnras/stw667}, \href
  {https://ui.adsabs.harvard.edu/abs/2016MNRAS.459..257S} {459, 257}

\bibitem[\protect\citeauthoryear{{Tamura} et~al.,}{{Tamura}
  et~al.}{2019}]{Tamura2019}
{Tamura} Y.,  et~al., 2019, \mndoi [\apj] {10.3847/1538-4357/ab0374}, \href
  {https://ui.adsabs.harvard.edu/abs/2019ApJ...874...27T} {874, 27}

\bibitem[\protect\citeauthoryear{{Toshikawa} et~al.,}{{Toshikawa}
  et~al.}{2018}]{Toshikawa2018}
{Toshikawa} J.,  et~al., 2018, \mndoi [\pasj] {10.1093/pasj/psx102}, \href
  {https://ui.adsabs.harvard.edu/abs/2018PASJ...70S..12T} {70, S12}

\bibitem[\protect\citeauthoryear{{Umehata} et~al.,}{{Umehata}
  et~al.}{2019}]{Umehata2019}
{Umehata} H.,  et~al., 2019, \mndoi [Science] {10.1126/science.aaw5949}, \href
  {https://ui.adsabs.harvard.edu/abs/2019Sci...366...97U} {366, 97}

\bibitem[\protect\citeauthoryear{{Valiante} et~al.,}{{Valiante}
  et~al.}{2016}]{Valiante2016}
{Valiante} E.,  et~al., 2016, \mndoi [\mnras] {10.1093/mnras/stw1806}, \href
  {https://ui.adsabs.harvard.edu/abs/2016MNRAS.462.3146V} {462, 3146}

\bibitem[\protect\citeauthoryear{{Venemans} et~al.,}{{Venemans}
  et~al.}{2007}]{Venemans2007}
{Venemans} B.~P.,  et~al., 2007, \mndoi [\aap] {10.1051/0004-6361:20053941},
  \href {https://ui.adsabs.harvard.edu/abs/2007A&A...461..823V} {461, 823}

\bibitem[\protect\citeauthoryear{{Vieira} et~al.,}{{Vieira}
  et~al.}{2013}]{Vieira2013}
{Vieira} J.~D.,  et~al., 2013, \mndoi [\nat] {10.1038/nature12001}, \href
  {https://ui.adsabs.harvard.edu/abs/2013Natur.495..344V} {495, 344}

\bibitem[\protect\citeauthoryear{{Viero} et~al.,}{{Viero}
  et~al.}{2014}]{Viero2014}
{Viero} M.~P.,  et~al., 2014, \mndoi [\apjs] {10.1088/0067-0049/210/2/22},
  \href {https://ui.adsabs.harvard.edu/abs/2014ApJS..210...22V} {210, 22}

\bibitem[\protect\citeauthoryear{{Wagner}, {Courteau}  \& {Brodwin}}{{Wagner}
  et~al.}{2017}]{Wagner2017}
{Wagner} C.~R.,  {Courteau} S.,   {Brodwin} M. e.~a.,  2017, \mndoi [\apj]
  {10.3847/1538-4357/834/1/53}, \href
  {http://adsabs.harvard.edu/abs/2017ApJ...834...53W} {834, 53}

\bibitem[\protect\citeauthoryear{{Walter} et~al.,}{{Walter}
  et~al.}{2012}]{Walter2012}
{Walter} F.,  et~al., 2012, \mndoi [\nat] {10.1038/nature11073}, \href
  {https://ui.adsabs.harvard.edu/abs/2012Natur.486..233W} {486, 233}

\bibitem[\protect\citeauthoryear{{Wang} et~al.,}{{Wang}
  et~al.}{2011}]{Wang2011}
{Wang} J.,  et~al., 2011, \mndoi [\mnras] {10.1111/j.1365-2966.2011.18220.x},
  \href {https://ui.adsabs.harvard.edu/abs/2011MNRAS.413.1373W} {413, 1373}

\bibitem[\protect\citeauthoryear{{Wang} et~al.,}{{Wang}
  et~al.}{2016}]{Wang2016}
{Wang} T.,  et~al., 2016, \mndoi [\apj] {10.3847/0004-637X/828/1/56}, \href
  {http://adsabs.harvard.edu/abs/2016ApJ...828...56W} {828, 56}

\bibitem[\protect\citeauthoryear{{Wang} et~al.,}{{Wang}
  et~al.}{2019a}]{Wang2019}
{Wang} T.,  et~al., 2019a, \mndoi [\nat] {10.1038/s41586-019-1452-4}, \href
  {https://ui.adsabs.harvard.edu/abs/2019Natur.572..211W} {572, 211}

\bibitem[\protect\citeauthoryear{{Wang}, {Pearson}, {Cowley}, {Trayford},
  {B{\'e}thermin}, {Gruppioni}, {Hurley}  \& {Micha{\l}owski}}{{Wang}
  et~al.}{2019b}]{WangL2019}
{Wang} L.,  {Pearson} W.~J.,  {Cowley} W.,  {Trayford} J.~W.,  {B{\'e}thermin}
  M.,  {Gruppioni} C.,  {Hurley} P.,   {Micha{\l}owski} M.~J.,  2019b, \mndoi
  [\aap] {10.1051/0004-6361/201834093}, \href
  {https://ui.adsabs.harvard.edu/abs/2019A&A...624A..98W} {624, A98}

\bibitem[\protect\citeauthoryear{{Wardlow} et~al.,}{{Wardlow}
  et~al.}{2013}]{Wardlow2013}
{Wardlow} J.~L.,  et~al., 2013, \mndoi [\apj] {10.1088/0004-637X/762/1/59},
  \href {https://ui.adsabs.harvard.edu/abs/2013ApJ...762...59W} {762, 59}

\bibitem[\protect\citeauthoryear{{Wei{\ss}} et~al.,}{{Wei{\ss}}
  et~al.}{2013}]{Weiss2013}
{Wei{\ss}} A.,  et~al., 2013, \mndoi [\apj] {10.1088/0004-637X/767/1/88}, \href
  {https://ui.adsabs.harvard.edu/abs/2013ApJ...767...88W} {767, 88}

\bibitem[\protect\citeauthoryear{{Wen} \& {Han}}{{Wen} \&
  {Han}}{2018}]{WenHan2018}
{Wen} Z.~L.,  {Han} J.~L.,  2018, \mndoi [\mnras] {10.1093/mnras/sty2533},
  \href {https://ui.adsabs.harvard.edu/abs/2018MNRAS.481.4158W} {481, 4158}

\bibitem[\protect\citeauthoryear{{Williams} et~al.,}{{Williams}
  et~al.}{2019}]{Williams2019}
{Williams} C.~C.,  et~al., 2019, \mndoi [\apj] {10.3847/1538-4357/ab44aa},
  \href {https://ui.adsabs.harvard.edu/abs/2019ApJ...884..154W} {884, 154}

\bibitem[\protect\citeauthoryear{{Wright} et~al.,}{{Wright}
  et~al.}{2010}]{Wright2010}
{Wright} E.~L.,  et~al., 2010, \mndoi [\aj] {10.1088/0004-6256/140/6/1868},
  \href {https://ui.adsabs.harvard.edu/abs/2010AJ....140.1868W} {140, 1868}

\bibitem[\protect\citeauthoryear{{Yang} et~al.,}{{Yang}
  et~al.}{2017}]{Yang2017}
{Yang} C.,  et~al., 2017, \mndoi [\aap] {10.1051/0004-6361/201731391}, \href
  {https://ui.adsabs.harvard.edu/abs/2017A&A...608A.144Y} {608, A144}

\bibitem[\protect\citeauthoryear{{Younger} et~al.,}{{Younger}
  et~al.}{2007}]{Younger2007}
{Younger} J.~D.,  et~al., 2007, \mndoi [\apj] {10.1086/522776}, \href
  {https://ui.adsabs.harvard.edu/abs/2007ApJ...671.1531Y} {671, 1531}

\bibitem[\protect\citeauthoryear{{Zhang} et~al.,}{{Zhang}
  et~al.}{2018a}]{ZhangZ2018}
{Zhang} Z.-Y.,  et~al., 2018a, \mndoi [\mnras] {10.1093/mnras/sty2082}, \href
  {https://ui.adsabs.harvard.edu/abs/2018MNRAS.481...59Z} {481, 59}

\bibitem[\protect\citeauthoryear{{Zhang}, {Romano}, {Ivison}, {Papadopoulos}
  \& {Matteucci}}{{Zhang} et~al.}{2018b}]{Zhang2018}
{Zhang} Z.-Y.,  {Romano} D.,  {Ivison} R.~J.,  {Papadopoulos} P.~P.,
  {Matteucci} F.,  2018b, \mndoi [\nat] {10.1038/s41586-018-0196-x}, \href
  {https://ui.adsabs.harvard.edu/abs/2018Natur.558..260Z} {558, 260}

\bibitem[\protect\citeauthoryear{{de Haan} et~al.,}{{de Haan}
  et~al.}{2016}]{deHaan2016}
{de Haan} T.,  et~al., 2016, \mndoi [\apj] {10.3847/0004-637X/832/1/95}, \href
  {https://ui.adsabs.harvard.edu/abs/2016ApJ...832...95D} {832, 95}

\makeatother
\end{thebibliography}

\end{document}